\documentclass{jfm}
\usepackage{graphicx}
\usepackage{multicol}
\usepackage{subcaption}
\usepackage{tabularx}
\usepackage{epstopdf, epsfig}
\usepackage{amsmath}
\usepackage{listings}
\usepackage[hidelinks]{hyperref}

\usepackage{xcolor}

\usepackage{soul}
\graphicspath{{Images/}}

\newcommand{\wx}[1][n,q]{w^{[#1]}_x}

\newcommand{\hvx}[1][n,q]{\hat{\gamma}^{[#1]}_x}

\newcommand{\dRod}[1][n,q]{\partial\mathcal{C}^{[#1]}}

\newcommand{\chij}[1][j]{\chi_{#1}}

\newcommand{\FT}[1]{\mathcal{F}\left[ #1 \right]}
\newcommand{\FTinv}[1]{\mathcal{F}^{-1}\left[ #1 \right]}

\newcommand{\cinxy}{\circ\in\{x,y\}}

\title{Seismic response of cylinder assemblies in axial flow}
\shorttitle{Seismic response of assemblies}

\author{Roberto Capanna\aff{1}
  \corresp{\email{capanna@gwu.edu}},
  Guillaume Ricciardi\aff{1},
  Emmanuelle Sarrouy\aff{2}
 \and Christophe Eloy\aff{3}}
\shortauthor{R. Capanna, G. Ricciardi, E. Sarrouy and C. Eloy}

\affiliation{\aff{1} CEA, DES, IRESNE, Department of Nuclear Technology, Cadarache, 13108, Saint-Paul-Lez-Durance, France
\aff{2}Aix Marseille Univ, CNRS, Centrale Marseille, LMA UMR 7031, Marseille, France
\aff{3}Aix Marseille Univ, CNRS, Centrale Marseille, IRPHE, Marseille, France}

\begin{document}

\maketitle

\begin{abstract}
Earthquakes are a great challenge for the safety of nuclear reactors. To address this challenge, we need to better understand how the reactor core responds to seismic forcing. 
The reactor core is made of fuel assemblies, which are themselves composed of flexible fuel rods immersed in a strong axial flow. This gives rise to strongly-coupled fluid-structure interactions whose accurate modelling generally requires high computational costs.
In this paper, we introduce a new model able to capture the mechanical response of the reactor core subjected to seismic forcing with low computational costs. 
This model is based on potential flow theory for the fluid part and Euler-Bernoulli beam theory for the structural part allowing us to predict the response to seismic forcing in presence of axial flow.. The linear equations are solved in the Fourier space to decrease computational time. 
For validation purposes, we first use the proposed model to compute the response of a single cylinder in axial flow. 
We then implement a multiple cylinder geometry made of 4 fuel assemblies, each made of $8\times 8$ cylinders, corresponding to an  experimental facility available at CEA. The comparison between numerical results and experiments show good agreement. The model can correctly predict the added mass. It can also qualitatively capture the coupling between assemblies and the effect of confinement. 
This shows that a potential flow approach can give insight into the complex fluid-structure interactions within a nuclear reactor and, in particular, be used to predict the response to seismic forcing at low computational cost. 
\end{abstract}

%
\section{Introduction}
%
One of the main concerns for the safety of nuclear power plants is represented by earthquakes. 
During an earthquake, the main risk is that fuel assemblies start to move and potentially touch each other or prevent the drop of the control rods used to cool the core. 
To better understand the motion of fuel assemblies subjected to seismic forcing, fluid-structure interaction models are needed.  In this paper, we present such a model, which has been developed with the objective of gaining insight into the complex fluid-structure interactions at play inside a nuclear reactor, while remaining computationally effective.  

A reactor core of a pressurised water reactor (PWR) is made of fuel assemblies (between $150$ and $250$ depending on the power of the reactor). Each fuel assembly gathers about 100 fuel rods, stands between four and five meters high, has a cross section of about $20\times20\,$cm$^2$ and weights about $800\,$kg. The fuel rods, which contain the pellets of uranium, have a diameter of about $1\,$cm for $4\,$m in height; the space between each fuel rod in the assembly is about $3\,$mm. Fuel rods are held together by grids (between $8$ and $10$ depending on the power of the reactor) distributed along the height of the fuel assembly. Springs and dimples are used in between the grid and the rods to avoid any drop of the fuel rods. 

The fuel assemblies are cooled down by a strong axial flow. This water flow is upwards, with velocities of about $5\,$m$\,$s$^{-1}$ at $150\,$ bar and $310\,^{\circ}$C. The flow regime  is fully turbulent with a Reynolds number based on the rod diameter of $\Rey\approx 5\times 10^5$. Note that, even if the main flow is upwards, the root-mean-square-average of the transverse component is between 5 and 15\% of the vertical velocity. 

The presence of the water flow gives rise to strongly-coupled interactions between the fluid and the structure \citep{Chen_72}: the motion of the structure modifies the fluid flow, which itself exerts a force on the structure. A first attempt to describe fluid-structure interactions is to use the concepts of added mass and added damping. The added mass is the inertial mass added to a body because of the presence of a surrounding fluid. For simplicity it can be viewed as a volume of fluid moving with the same velocity as the body, though in reality all fluid particles will be moving to various degrees \citep{Lamb_1895,DuBuat}. The added mass $M_a$ of a non-deformable body moving at a velocity $U$ through an unbounded fluid otherwise at rest can be defined such that the kinetic energy in the volume of fluid is $E_k=\frac{1}{2}M_a U^2$ or equivalently
\begin{equation}
M_a=\frac{\rho}{U^2} \int \|\textbf{V}\|^2 \,dv,
\end{equation}
where $\textbf{V}$ is the velocity field.
A complete collection of added mass for different geometries and flow conditions can be found in \citet{Wendel_56} or \citet{Brennen_82}, for instance. 

While the added mass is mainly due to pressure forces exerted on the body, viscous forces and boundary layer separation give rise to drag and to an added damping effect. \citet{Taylor_52} proposed a model for the damping force on a slender structure, which has been widely used \citep[e.g.,][]{Paid_1966_Theo,Triantafyllou_85, Gosselin_2011, Singh_2012}. In Taylor's model,  the damping force is decomposed into two terms: a normal force akin to drag and a longitudinal force depending on the body surface and on the incident angle. 

Generally, the total fluid force exerted on the non-deformable body is assumed to be decomposed into its added-mass component (proportional to the acceleration) and its drag component (proportional to the square of the velocity). This decomposition, known as the Morison equation \citep{Morison}, can correctly describe experimental observations providing drag and inertial coefficients are  empirically adjusted (in general, they depend on the motion amplitude and frequency). 

For deformable cylinders in axial flow, the concept of added mass  does not apply directly. In that case, one can use  slender body theory developed by \citet{Lighthill_Slender_Body, Lighthill_1960}. This theory makes use of potential flow theory and arguments of momentum balance in slices of fluids along the slender body. The resulting normal force per unit length exerted by the fluid  can be written in the limit of small displacement
\begin{equation}
F(X,T)=- M_a\left(\partial_T + U\partial_X\right)^2 W,
\label{Eq::Lighthill_3}
\end{equation}
where $W(X,T)$ is the normal displacement, $U$ is the undisturbed axial flow velocity, and here $M_a$ is the added mass per unit length ($M_a=\rho\pi A^2$, for a circular cylinder of radius $A$). In the case of a solid body, $W$ has no $X$-dependence and one recovers a force opposite to $M_a\partial^2_T W$, as expected. 

Based on the work of \citet{Lighthill_1960}, \citet{Paid_1966_Theo,Paid_1966_Exp} studied theoretically and experimentally the dynamics of a flexible slender cylinder clamped at its leading edge, which is immersed in an axial flow. He found that, above a critical flow velocity, a flutter instability appears. Later, \citet{Paid_1973} and \citet{Paid_1979} extended this flutter stability analysis to confined geometries, by using the work of \citet{Clasen_72} and \citet{Chen_72} on added mass in confined geometries. 

The stability analysis of a single cylinder in axial flow has been later generalized to clusters of cylinders \citep{Paid_1973,Paid_1977_a,Paid_1977_b, Paid_1979_Clusters}. If each cylinder is free to move independently, the fluid acts as a coupling medium and the motions of the cylinders are synchronized. Above a critical flow velocity, the coupled pinned-pinned cylinders loose stability by divergence (unstable mode of eigenfrequency zero). 
For higher flow velocities, the system may be subjected to several divergence and flutter instabilities simultaneously. In  recent years, this work of Pa\"idoussis has been extended \citep[e.g.,][]{De_Langre,Schouveiler_2009,Michelin_2009} and studied numerically both for a single cylinder \citep{De_Ridder_13,De_Ridder_15} and for a cluster of cylinders \citep{De_Ridder_17}.  

The flutter of a flexible cylinder in axial flow bears similarities with the flutter of an elastic plate often referred to as the flag instability. For a plate, two asymptotic limits can be studied: slender structures where Lighthill's slender body theory applies, and wide plates for which a two-dimensional approach is suited \citep{Wu_2001}. For intermediate aspect ratios, the pressure distribution on a flexible plate can be solved by projecting the problem in Fourier space \citep{Guo_2000,Eloy_2007, Eloy_2008, Doare_2, Doare_3,Doare_2011}. With this projection, the approaches of slender body approximation and large span approximation can be generalised to any aspect ratio. 
In this paper, we will follow a similar approach to describe the flow around assemblies of flexible cylinders found in PWR.

A different approach was proposed by \citet{Ricciardi_2009} using a porous medium approach. With this approach, it  becomes possible to model both the fluid and the structure dynamics of a whole core. Some local information are lost compared to a direct numerical simulation, such as the vibrations of individual rods, but fluid-mediated interactions between fuel assemblies can be modelled. This model shows good agreement with experimental results on the response of the whole core to external forcing, but computational cost remains high.

At present, the models describing the fluid-structure interactions within a reactor core can be divided into two families: (1) complex numerical models with high computational costs, which usually hinder the physical mechanisms; and (2) linear models with low computational costs, which do not capture important mechanisms such as fluid-mediated interactions between assemblies. 
In this paper, we introduce a new model based  on potential flow theory, with the objective of providing an accurate, computationally effective modelling of fluid-structure interactions within a reactor core. 

The paper is organised as follows. In \S\ref{sec:Model}  the model is presented and the mathematical methodology used to solve the problem is described. The model is then applied to a single cylinder in  \S\ref{sec::SingleCylinder} and compared to results of the literature for validation purposes. In \S\ref{sss::MultipleCylinders_Assembly1}, it is applied to a multiple cylinder geometry, replicating the geometry of an experimental facility. Numerical results are compared to experimental data and discussed in \S\ref{sec:Experiments}. Finally, in \S\ref{sec:Conclusion}, some conclusions are drawn. 
%
\section{Potential flow model}\label{sec:Model}
%
In this section, we describe the proposed model where the fluid is treated as potential and the structure as an Euler-Bernoulli beam. The model is introduced for a single cylinder in axial flow, but as we shall see, it can be generalised to multiple cylinders. 
\subsection{Problem statement}
\begin{figure}
\centering
\begin{subfigure}{2.6in}
  \centering
  \includegraphics[width=2.2in]{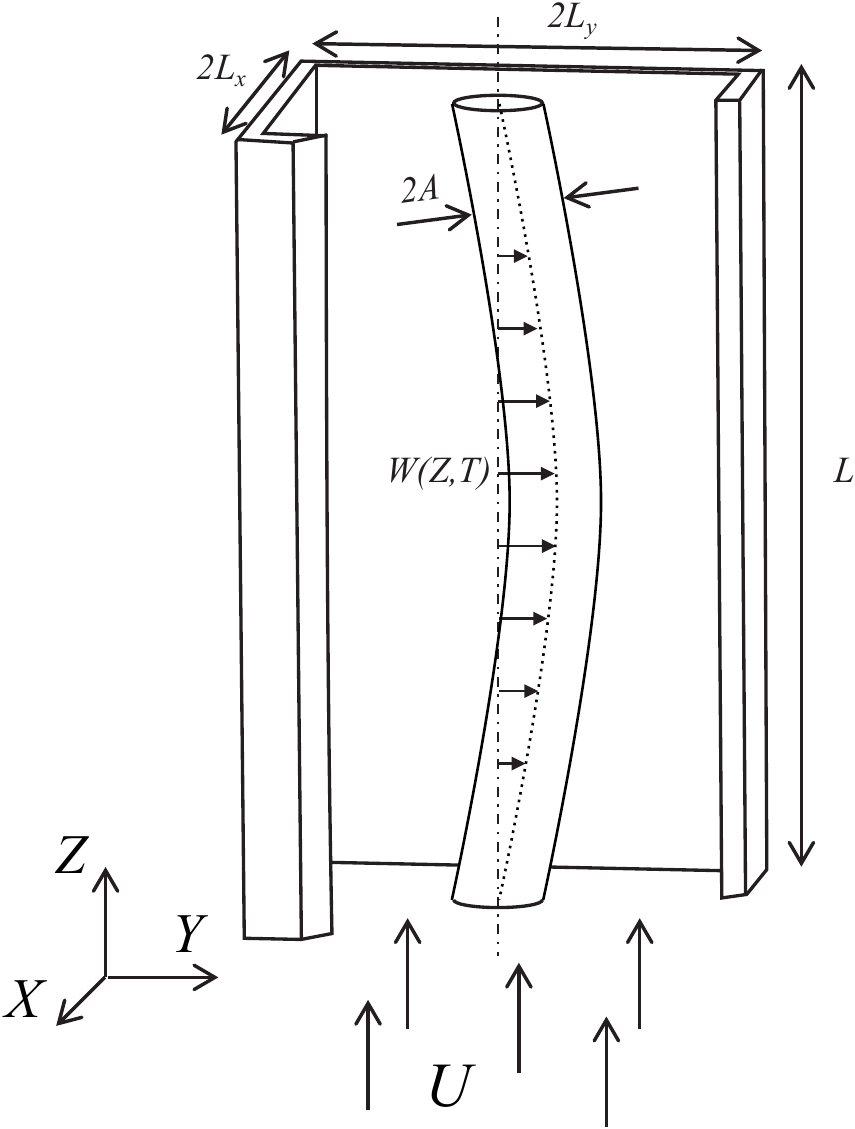}
  \caption{Real space.}
  \label{fig::Cylinder_3D}
\end{subfigure}
\begin{subfigure}{2.6in}
  \centering
  \includegraphics[width=2in]{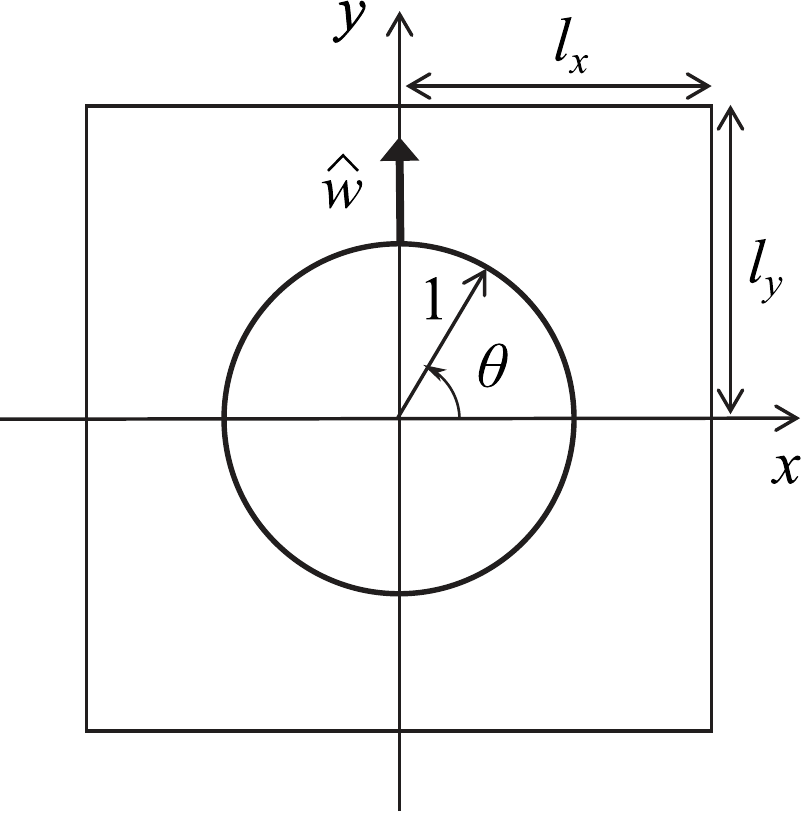}
  \caption{Fourier space.}
  \label{fig::Cylinder_2D}
\end{subfigure}
\caption{Representation of confined pinned-pinned cylinder deformed under axial flow in real space and in Fourier space.}
\label{fig::Cylinder_3D-2D}
\end{figure}

We consider a pinned-pinned cylinder of length $L$ and radius $A$ (Figure~\ref{fig::Cylinder_3D}). We will use two coordinate systems: a Cartesian system $(X,Y,Z)$ with its origin at the bottom of the cylinder and a cylindrical system $(R,\theta,Z)$ with same origin (Figure~\ref{fig::Cylinder_3D}).

The cylinder is immersed in a uniform axial flow with velocity $U$ and bounded in the $X$ and $Y$ directions by rigid walls at distances $L_{x}$ and $L_{y}$  from the cylinder axis. Without loss of generality, we consider that the cylinder deflects in the $Y$-direction and the deflection is called $W(Z,T)$ (Figure~\ref{fig::Cylinder_3D}). 

We use the linearised Euler-Bernoulli beam equation \citep{Bauchau} to describe the dynamics of the cylinder deflection
\begin{equation}
M_s\partial^2_T W + B\partial^4_Z W=F_Y,
\label{Eq::Euler_Bernoulli}
\end{equation}
where $M_s$ is the mass of the cylinder per unit length, $B=EI$ is the bending rigidity ($E$ being the Young's modulus and $I=\frac{1}{4}\pi A^4$ the second moment of area), and $F_Y$ is the force per unit length that the fluid exerts on the cylinder in the $Y$-direction. 

We assume that the forces exerted by the fluid on the structure are mainly originating from pressure difference. Hence, the force per unit length can be expressed as 
\begin{equation}
F_Y(Z,T)=-\int_{0}^{2\pi}P(A,\theta,Z,T)\sin(\theta)\,A d \theta,
\label{Eq::Force_Pressure}
\end{equation}
where $P(R,\theta,Z,T)$ is the pressure field in cylindrical coordinates. 

The problem is made dimensionless using the cylinder radius $A$, the flow velocity $U$ and the fluid density $\rho$ as characteristic length, speed and density. Dimensionless quantities are designated with lowercase letters such that, for instance,
\begin{eqnarray*}
r=\frac{R}{A},
\quad
w=\frac{W}{A},
\quad
z=\frac{Z}{A},
\quad
l_x=\frac{L_x}{A},
\quad
l_y=\frac{L_y}{A},
\quad
t=\frac{UT}{A},
\quad
p=\frac{P}{\rho U^2}.
\end{eqnarray*}
In dimensionless form, the linearised Euler-Bernoulli equation (\ref{Eq::Euler_Bernoulli}) becomes
\begin{equation}
m\,\partial^2_t w + b \, \partial^4_z w = f_y,
\label{Eq::Euler_Bernoulli_Adm}
\end{equation}
where 
\begin{eqnarray*}
m=\frac{M_s}{\rho A^2}, \qquad
b=\frac{B}{\rho U^2 A^4}, \qquad
f_y=\frac{F_Y}{\rho U^2 A}.
\end{eqnarray*}

The flow is assumed to be potential, inviscid and incompressible, such that the flow velocity is given by 
\begin{equation}
\textbf{V} = U \textbf{e}_z + \nabla \Phi,
\end{equation}
where $\textbf{e}_z$ is the unit vector along $z$ and $\Phi(X,Y,Z,T)$ is the perturbation potential. In dimensionless units, this potential is $\phi=\Phi/(UA)$. 
To find the flow around the moving cylinder, one has to solve a Laplace problem with Neumann boundary conditions (linearised for small displacements)
\begin{eqnarray}
\Delta {\phi} &=&0 ,  \label{Eq::Bound_Cond_Poisson_Prob0}\\
\left.\frac{\partial {\phi}}{\partial {x}}\right|_{|{x}|={l}_x} &=&0 , \label{Eq::Bound_Cond_Poisson_Proba}\\
\left.\frac{\partial {\phi}}{\partial {y}}\right|_{|{y}|={l}_y} &=&0 , \label{Eq::Bound_Cond_Poisson_Probb}\\
\left.\frac{\partial {\phi}}{\partial {n}}\right|_{{r}=1} &=& -(\partial_{{t}}+\partial_{{z}}){w}(z,t)\sin\theta ,
\label{Eq::Bound_Cond_Poisson_Prob}
\end{eqnarray}
where (\ref{Eq::Bound_Cond_Poisson_Proba}) and (\ref{Eq::Bound_Cond_Poisson_Probb}) come from the impermeability of the walls and (\ref{Eq::Bound_Cond_Poisson_Prob}) from the impermeability on the cylinder wall.

Using the linearised unsteady Bernoulli equation, the dimensionless pressure field can be linked to $\phi$ \citep{Capanna_PhD}
\begin{equation}
{p}(x,y,z,t)=-(\partial_{{t}}+\partial_{{z}}){\phi}.
\label{Eq::Bernoulli_LinAdim}
\end{equation}
Using this relation and applying the operator $(\partial_{{t}}+\partial_{{z}})$ to the system (\ref{Eq::Bound_Cond_Poisson_Prob0}--\ref{Eq::Bound_Cond_Poisson_Prob}) above yields
\begin{eqnarray}
\Delta {p} &=&0 , 
\label{Eq::Laplace_Adim1} \\
\left.\frac{\partial {p}}{\partial {x}}\right|_{|{x}|={l}_x} &=&0 ,
\label{Eq::Laplace_Adim2} \\
\left.\frac{\partial {p}}{\partial {y}}\right|_{|{y}|={l}_y} &=&0 ,
\label{Eq::Laplace_Adim3} \\
\left.\frac{\partial {p}}{\partial {n}}\right|_{{r}=1} &=& (\partial_{{t}}+\partial_{{z}})^2{w}({z},{t})\sin\theta .
\label{Eq::Laplace_Adim4}
\end{eqnarray}
It shows that the pressure field is also a solution to a Laplace equation with Neumann boundary conditions. This is why the term $P/\rho$ is called the acceleration potential in airfoil theory.   
\subsection{Problem in Fourier space}
To solve the set of equations (\ref{Eq::Laplace_Adim1}--\ref{Eq::Laplace_Adim4}), we will use the method proposed by \cite{Doare_2011}. It consists in expressing the Laplace problem in Fourier space along $z$
\begin{equation}
\hat{p}(x,y,k,t)=\FT{p(x,y,z,t)},
\label{Eq::Pressure_Fourier_Transf}
\end{equation}
where $\hat{p}$ is the Fourier transform of $p$ with $\FT{\cdot}$ defined as follows
\begin{equation}\label{eq:FT_Definition}
\FT{f(z)}=\frac{1}{2\pi}\int_{-\infty}^{+\infty}f(z)e^{-ikz} \, dz = \hat{f}(k),
\end{equation}
together with the inverse Fourier transform
\begin{equation}
    \FTinv{\hat{f}(k)} = \int_{-\infty}^{+\infty}\hat{f}(k)e^{ikz} \, dk = f(z).
\end{equation}
The convolution product along $z$, noted $\star$, is also introduced 
\begin{equation}\label{eq.convolution}
    f\star g = \int_{-\infty}^{+\infty}f(\zeta)g(z-\zeta)\,d\zeta = 2 \pi \FTinv{\hat{f}\,\hat{g}}.
\end{equation}

In Fourier space, the three-dimensional Laplace problem is transformed into a two-dimensional Helmholtz problem
\begin{eqnarray}
(\partial_{{x}}^2+\partial_{{y}}^2)\hat{p} &=& {k}^2\hat{p} , \label{Eq::Laplace_Fourier_1}\\
\left.\frac{\partial \hat{p}}{\partial {x}}\right|_{|{x}|={l}_x} &=&0 ,\label{Eq::Laplace_Fourier_1x}\\
\left.\frac{\partial \hat{p}}{\partial {y}}\right|_{|{y}|={l}_y} &=&0 ,\label{Eq::Laplace_Fourier_1y}\\
\left.\frac{\partial \hat{p}}{\partial {n}}\right|_{{r}=1} &=&\hat{\gamma}({k},{t})\sin\theta , \label{Eq::Laplace_Fourier_2}
\end{eqnarray}
where $\hat{\gamma}(k,t)$ is the Fourier transform of the impermeability boundary condition, such that
\begin{equation}
\hat{\gamma} = \FT{(\partial_{{t}}+\partial_{{z}})^2{w}}.
\label{Eq::Boundary_Fourier}
\end{equation}

Equation (\ref{Eq::Force_Pressure}) relates the force per unit length exerted by the fluid on the cylinder and the pressure field. Using dimensionless quantities and taking the Fourier transform of (\ref{Eq::Force_Pressure}), one obtains the equivalent in Fourier space
\begin{equation}
\hat{f}_y(k,t) = -\int_0^{2\pi} \hat{p} (r=1,\theta,k,t) \sin\theta\,d\theta,
\label{eq:hat_f_y}
\end{equation}
where $\hat{f}_y$ is the Fourier transform of $f_y$. 
The pressure $\hat{p}$ being solution of the linear Helmholtz problem (\ref{Eq::Laplace_Fourier_1}--\ref{Eq::Laplace_Fourier_2}), its solution is proportional to $\hat{\gamma}(k, t)$. Hence, (\ref{eq:hat_f_y}) can be written
\begin{equation}
\hat{f}_y = -\hat{\mu}(k) \hat{\gamma}(k,t). \label{Eq::Fourier_fy}
\end{equation}
where the function $\hat{\mu}(k)$ depends on $l_x$ and $l_y$ through the boundary conditions (\ref{Eq::Laplace_Fourier_1x}) and (\ref{Eq::Laplace_Fourier_1y}).

Equation (\ref{Eq::Euler_Bernoulli_Adm}) governing beam dynamics then becomes in Fourier space
\begin{equation}\label{beam_eq_Fourier}
    m\partial^2_t\hat{w} + k^4 b \hat{w} = \hat{\mu}(k)\left[-(\partial^2_t\hat{w}+2ik\partial_t\hat{w}-k^2\hat{w})\right],
\end{equation}
showing that $\hat{\mu}$ plays the role of an added mass in Fourier space and that added damping $2ik\hat{\mu}$ and added stiffness $-k^2\hat{\mu}$ are proportional to this added mass.

The objective now will be to calculate the added mass $\hat{\mu}(k)$. In practice, to calculate it, we can assume $\hat{\gamma} = 1$, compute numerically the solution $\hat{p}$ of the Helmholtz problem (\ref{Eq::Laplace_Fourier_1}--\ref{Eq::Laplace_Fourier_2}) for different values of $k$, $l_x$ and $l_y$, and finally use \eqref{eq:hat_f_y} and \eqref{Eq::Fourier_fy} to compute $\hat{\mu}(k)$.

\subsection{Modal decomposition and range of interest for k}
Considering an external forcing at a given frequency $\Omega$, the displacement of the cylinder can be written in dimensionless form as
\begin{equation}\label{eq.modal_decomposition}
    w(z,t) = \sum_{j=1}^{\infty}q_j(t)\chij(z),
\end{equation}
where $\chij(z)$ are the beam eigenmodes and $q_j(t) = \eta_j e^{i\omega t} $ the generalised coordinates, with $\eta_j$ the modal amplitudes (which do not depend on time), and $\omega=\Omega A/U$ the dimensionless forcing frequency.

\begin{figure}
\centering
\includegraphics[width=.99\textwidth]{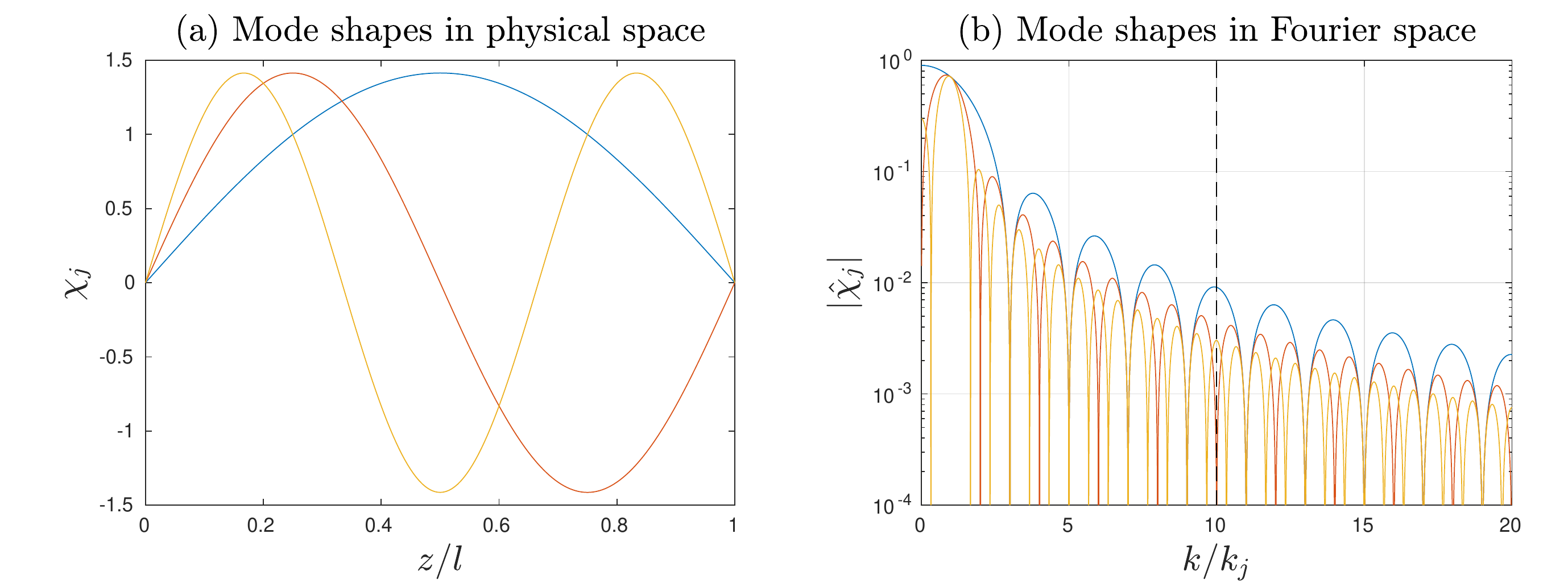}
\caption{(a) Eigenmodes for the first 3 modes of a pinned-pinned beam and (b) their Fourier transform.}
\label{fig::Modal_Shapes}
\end{figure}

To determine the interesting range for $k$ for which $\hat{\mu}$ should be computed, let us examine the eigenmodes $\chij(z)$ and their Fourier transform $\hat{\chi}_j(k)$ for pinned-pinned beams.
Eigenmodes are given by \cite{Blevins}
\begin{equation}\label{eq.eigenmodes}
    \chij(z) = \sqrt{2}\sin(k_jz),\quad
    \mbox{with } k_j=j\pi/l,
\end{equation}
with $l=L/A$ the dimensionless cylinder length. Using this normalisation, the scalar product of two eigenmodes is 
\begin{equation}\label{eq:Modes_Normalisation}
    \langle\chi_i,\chi_j\rangle=\frac{1}{l}\int_0^l\chi_i(z)\chi_j(z)\,dz = \delta_{ij}.
\end{equation}
As Figure~\ref{fig::Modal_Shapes}(b) shows, $k$ values greater than $10k_j$ have contributions at least 100 times smaller than low $k$ values in Fourier space. Hence, they can be truncated.

Classical values for PWR beams are $A\in[10^{-3},10^{-2}]$\,m and $L\in[1,10]$\,m. Hence, classical slenderness ratios $l$ belong to $[10^2,10^4]$ and dimensionless wavenumbers $k_j=j\pi/l\in[3\times 10^{-4}, 0.3]$ when considering the first 10 modes. This implies that the interesting range for which $\hat{\mu}$ should be evaluated is roughly $k\in[10^{-4}, 10]$.

\subsection{Equation of motion in generalised coordinates}\label{sec.generalised_coordinates}
Taking the Fourier transform of the modal decomposition \eqref{eq.modal_decomposition}, inserting it into the beam equation \eqref{beam_eq_Fourier} in Fourier space, performing an inverse Fourier transform, and taking the scalar product \eqref{eq:Modes_Normalisation} with the eigenmodes $\chij$ gives a set of differential equations for the generalised coordinates $q_i(t)$ that can be written in vector form as
\begin{equation}\label{eq.generalised_coordinates}
\mathsfbi{M}\, \mathbf{\ddot{q}} + \mathsfbi{K}\, \mathbf{q} = 
    -\left(  
        \mathsfbi{M_a}\, \mathbf{\ddot{q}} + \mathsfbi{C_a}\, \mathbf{\dot{q}} + \mathsfbi{K_a}\, \mathbf{q}
    \right),
\end{equation}
where $\mathbf{q}(t) = [q_1(t), q_2(t), \cdots]^T$ is the vector of generalised coordinates, the left-hand side corresponds to the unforced beam equation, and the right-hand side corresponds to the forcing by pressure forces. Here, we have omitted the external forcing term representing the seismic forcing.

The matrix $\mathsfbi{M} = m \mathsfbi{I}$, with $\mathsfbi{I}$ the identity matrix, is the mass matrix, and $\mathsfbi{K} = b\, \mathsfbi{diag}(\{k_j^4\})$ is the stiffness matrix (where $\mathsfbi{diag}(\{k_j^4\})$ means the diagonal matrix with $k_1^4$, $k_2^4$, etc. on the diagonal). Equating the left-hand side of \eqref{eq.generalised_coordinates} to zero allows to recover the eigenmodes and the eigenfrequencies $\omega_j = \sqrt{bk_j^4/m}$ of the beam in vacuum. 

The matrices $\mathsfbi{M_a}$, $\mathsfbi{C_a}$, and $\mathsfbi{K_a}$ correspond to the added mass, added damping and added stiffness matrices respectively. Their coefficients can be calculated as 
\begin{subequations}\label{eq.added_matrices}
\begin{eqnarray}
    \left(\mathsfbi{M_a}\right)_{ij} & = & 
        \langle \chi_i, \FTinv{\hat\mu \hat{\chi}_j} \rangle =
        \frac{1}{2\pi} \langle \chi_i, \mu \star \chi_j \rangle, \\
    \left(\mathsfbi{C_a}\right)_{ij} & = & 
        \langle \chi_i, \FTinv{2ik\hat\mu \hat{\chi}_j} \rangle =
        \frac{1}{2\pi} \langle \chi_i, 2 \mu \star \chi'_j \rangle, \\
    \left(\mathsfbi{K_a}\right)_{ij} & = & 
        \langle \chi_i, \FTinv{-k^2\hat\mu \hat{\chi}_j} \rangle =
        \frac{1}{2\pi} \langle \chi_i, \mu \star \chi''_j \rangle,
\end{eqnarray}
\end{subequations}
where $\chi_j'$ and $\chi_j''$ respectively denote first and second derivatives of $\chi_j$, $\mu = \FTinv{\hat\mu}$ and  we have used the property \eqref{eq.convolution} of the convolution product. 

We see here that the knowledge of $\hat\mu(k)$ or $\mu(z)$ is enough to compute the matrices $\mathsfbi{M_a}$, $\mathsfbi{C_a}$, and $\mathsfbi{K_a}$. In addition, these matrices only depend on the geometry of the problem (i.e. the confinements $l_x$ and $l_y$) and not on the forcing. Finally, note that the Helmholtz problem \eqref{Eq::Laplace_Fourier_1} only depends on $k^2$, which means that $\hat\mu(k)$ and $\mu(z)$ are real and even functions. The  coefficients of the matrices $\mathsfbi{M_a}$, $\mathsfbi{C_a}$, and $\mathsfbi{K_a}$ are thus all real, as expected. 

\section{Single cylinder}
\label{sec::SingleCylinder}
%
In this section, the numerical calculations performed for the single cylinder geometry using the code implemented on FreeFEM++ \citep{FreeFem_0, FreeFem_1} will be presented. FreeFEM++ requires the weak formulation of problems. Problem \eqref{Eq::Laplace_Fourier_1}-\eqref{Eq::Laplace_Fourier_2} is then implemented as follows
\begin{equation*}
\int_{\partial\mathcal{C}}v\,\sin(\theta)\,dS-\int_\mathcal{D}(\nabla\hat{p}\cdot \nabla v+k^2\hat{p}v)\,dV = 0, \forall v
\end{equation*}
where $\partial\mathcal{C}$ denotes the rod's border and $\mathcal{D}$ the fluid domain (see Appendix \ref{app:FreeFEM1Cylinder}).

\subsection{Meshing}
As previously described, the approach for the numerical resolution of the mathematical problem is to take advantage of the Fourier transform. This allows us to solve several 2D problems instead of solving a 3D problem. 

\begin{figure}
\centering
\def\svgwidth{.7\textwidth}\input{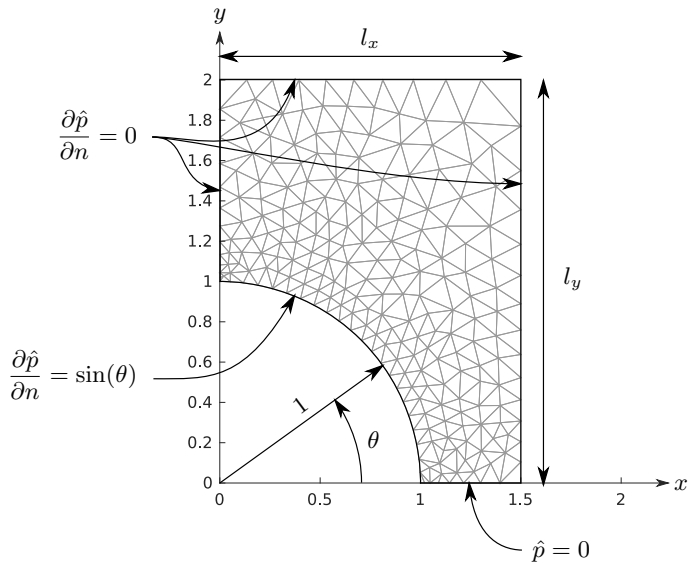}
\caption{Helmholtz problem represented in the FEM domain: limit conditions and example mesh for $k=0.01$, $l_x=1.5$ and $l_y=2$.}
\label{fig::FEM_Geometry}
\end{figure}

In this section, we present numerical solution of the Helmholtz problem described by Equations (\ref{Eq::Laplace_Fourier_1}-\ref{Eq::Laplace_Fourier_2}).  Our objective is to compute the value of the function $\hat{\mu}(k;l_x,l_y)$ for different values of $k$, $l_x$ and $l_y$. Thanks to the symmetries of the problem, the equations can be solved on a quarter of the domain only (Figure \ref{fig::FEM_Geometry}).  

Before showing the results of these numerical computations, some considerations have to be made on the discretisation of the domain. A convergence study has been performed in order to find the optimal meshing for different combinations of confinement size and wave number. Typically, the size of the meshes must be decreased when the wave number $k$ is increased. In addition, the refinement of meshes will increase as the confinement size decreases and when getting closer to the cylinder.

Explored values are $k\in[10^{-5},10^3]$, $l_x,l_y\in[1.25,50]$. The variations of the number of cells in the mesh with $l_x$ and $l_y$ do not depend on $k$: number of cells is maximum for $l_x=l_y=5$. This maximum is equal to 864 for $k\leq 1$ and increases quadratically for $k>1$. FreeFEM++ code is included in Appendix~\ref{app:FreeFEM1Cylinder}.

\subsection{Slender body limit}

In the asymptotic limit of long
wavelengths ($k\ll 1$) and negligible confinement ($l_x,l_y\gg 1$), we expect to recover the result of slender body theory \cite{Lighthill_1960} given by (\ref{Eq::Lighthill_3}), which writes in dimensionless form
\begin{equation}
f_y = - m_a (\partial_{{t}}+\partial_{{z}})^2 w,
\end{equation}
with $m_a = \pi$ the added mass per unit length of a circular cylinder in dimensionless units.
By analogy with (\ref{Eq::Fourier_fy}), one thus expects
\begin{equation}
\lim_{k\to 0} \hat{\mu} = \pi, \quad \mbox{for } l_x,l_y\gg 1. 
\end{equation}

\begin{figure}
\centering
\includegraphics[width=.99\textwidth]{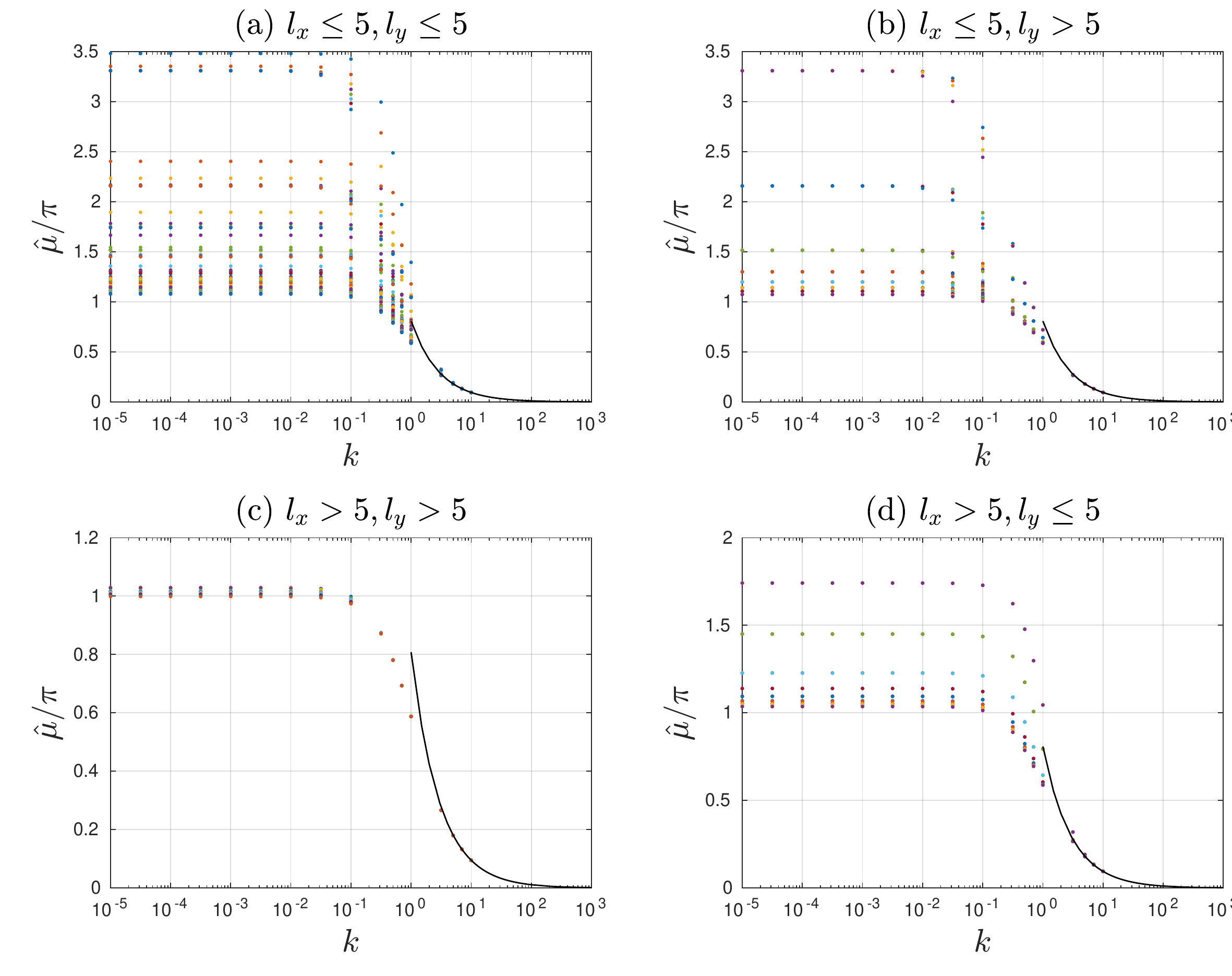}
\caption{Value of $\hat{\mu}$ as function of the wavenumber $k$ for several confinements ranges. Dots denote computed values and solid line for $k>1$ denotes the asymptotic extrapolation.}\label{fig::hatmuVSk}
\end{figure}

Figure \ref{fig::hatmuVSk} shows the behaviour of $\hat{\mu}$ as function of the wavenumber $k$ for different ranges of confinement sizes. The added mass does not change substantially for wavenumbers below $10^{-2}$ or above $10^{2}$, whatever the confinement. 
Moreover, for confinements larger than two diameters ($l_x,l_y>5$), the added mass converges towards $\pi$, as expected from slender body theory (Figure~\ref{fig::hatmuVSk}c). 

\subsection{Confinement Effects}\label{sse::SingleCylinder_Confinement}
\begin{figure}
    \centering
    \includegraphics[width=.9\textwidth]{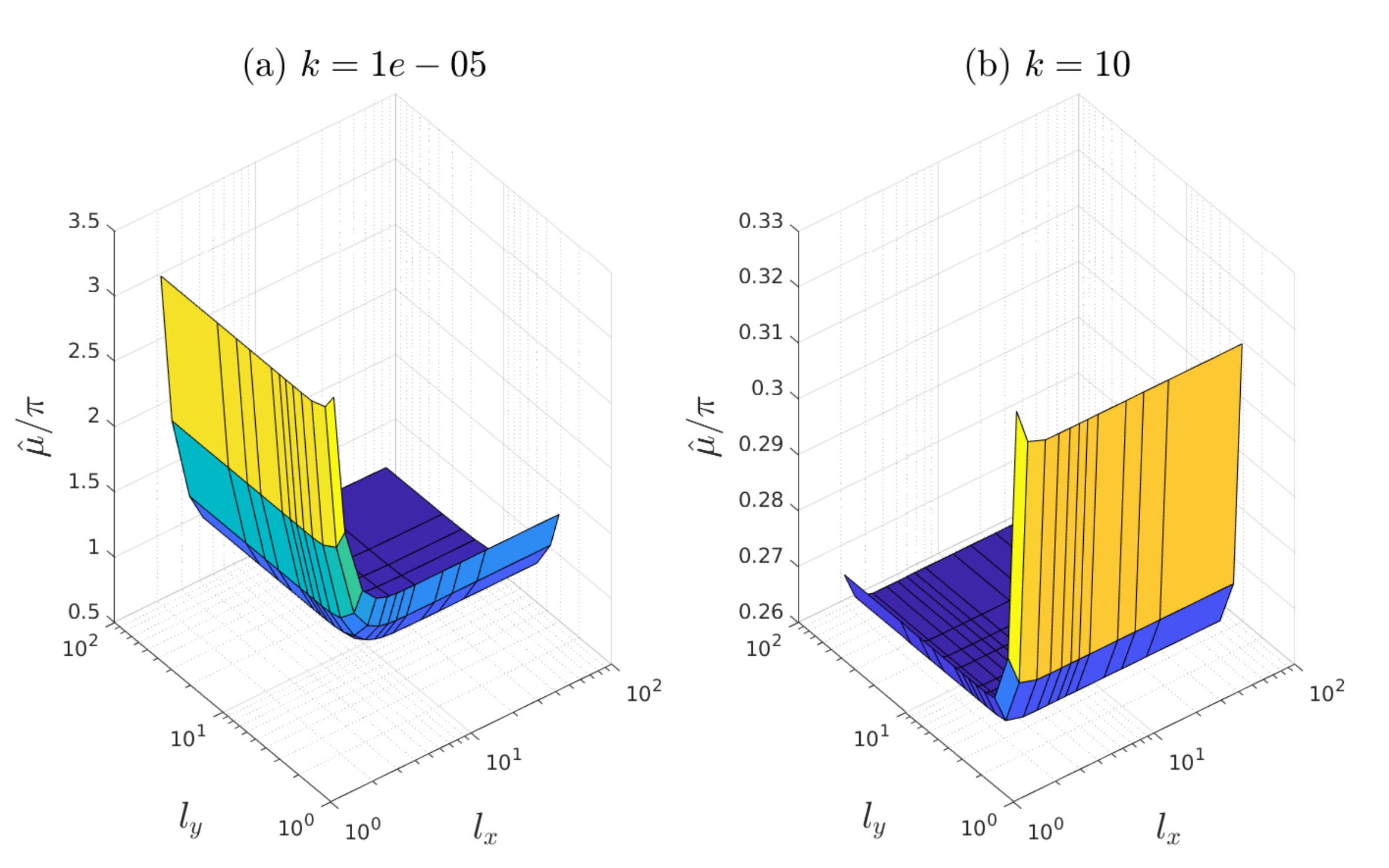}\\
    \caption{Confinement effects for small and big wavenumbers.}
    \label{fig::ConfinementEffect}
\end{figure}
Figure~\ref{fig::hatmuVSk} shows that $\hat{\mu}$ does not evolve much for small values of $k$, but still depends on the confinement size as soon as the unconfined hypothesis cannot be made (that is $l_x$ or $l_y <5$). 
Two extreme cases are considered: (1) very small values of wavenumbers ($k= 10^{-5}$); and (2) large wavenumbers ($k= 10$).

Figure \ref{fig::ConfinementEffect} shows how the added mass depend on both $x$ and $y$ confinement sizes for the two extreme wavenumbers considered ($k= 10^{-5}$ and $k=10$). For $k= 10^{-5}$, the confinement along the $x$-direction has much more influence than its counterpart along the $y$-direction (Figure~\ref{fig::ConfinementEffect}a). For $k=10$, this is the opposite. Note however that these effects have different magnitudes: for $k= 10^{-5}$, the value of $\hat{\mu}$ roughly triples between $l_x=10$ and $l_x=1.25$, while, for $k=10$, $\hat{\mu}$ varies only by 15\% between $l_y=10$ and $l_y=1.25$.

Figure~\ref{fig::hatmuVSk} thus shows that, for large wavenumbers, the dependency on the confinement is weak. Hence, we can assume that $\hat{\mu}$ is independent of $l_x$ and $l_y$, for large $k$ values. Based on this approximation, we can construct a power-law approximation of $\hat{\mu}$  displayed as black line in Figure~\ref{fig::hatmuVSk}
\begin{equation}
    \hat{\mu}\approx 2.5\,k^{-1},\ k\geq 10.
\end{equation}

\begin{figure}
    \centering
  \includegraphics[width=.9\textwidth]{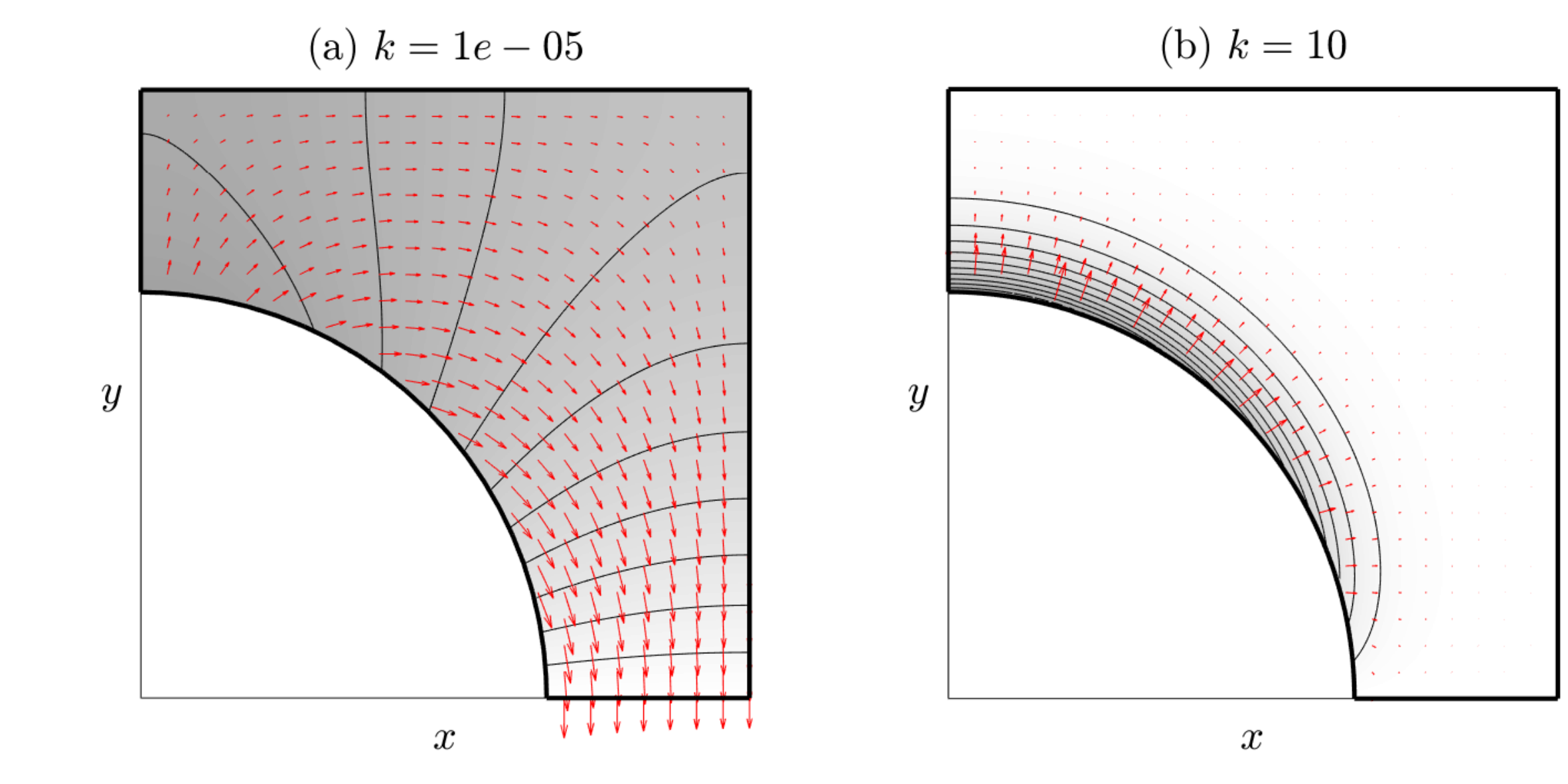}
    \caption{Velocity and pressure fields in Fourier space for both small and large wavenumbers.}
    \label{fig::pAndvFields}
\end{figure}

For small wavenumbers, although the cylinder can only move in the $y$ direction, it is the confinement along $x$ that has the strongest influence. To better understand this paradoxical result, we plot the velocity field in the Fourier space (Figure \ref{fig::pAndvFields}). For $k= 10^{-5}$, a strong flow in the $y$ direction (opposite to the cylinder displacement) is observed along the wall $x=l_x$ around the cylinder diameter ($y\approx0$) as showed in Figure \ref{fig::pAndvFields}a. For large wavenumbers however, no such flow is observed (Figure \ref{fig::pAndvFields}b). By virtue of mass conservation, when the confinement along $x$ is reduced, this flow increases. This is why there is a strong dependency of the fluid-structure interaction force on the $x$ direction confinements for small wavenumbers.

Finally, for all the wavenumbers that have been considered and for all confinements, the added mass $\hat{\mu}$ appears to increase as the confinement size decreases (in both directions). This observation is in agreement with the literature works on channel flows made by \cite{Chen_72, Chen_85, Paid_1979}. 

The calculations performed for a single cylinder geometry allowed us to prove the reliability of the proposed simplified model. The introduction of the potential flow theory and the use of Fourier transform are key to directly relate  the cylinder displacement  and the pressure resulting force. 
This approach leads to fast calculations for any kind of geometries.  In the next section, it will be applied to a group of cylinders and compared to experimental results in the same geometry.
%
\section{Assemblies of cylinders}
\label{sss::MultipleCylinders_Assembly1}
%
This section is dedicated to a geometry with multiple cylinders. This geometry is built to represent the four surrogate fuel bundles of the experimental facility ICARE that we used to validate our results.  

\subsection{ICARE experimental apparatus}\label{sse::ICARE}
\begin{figure}
\centering
\includegraphics[width=\textwidth]{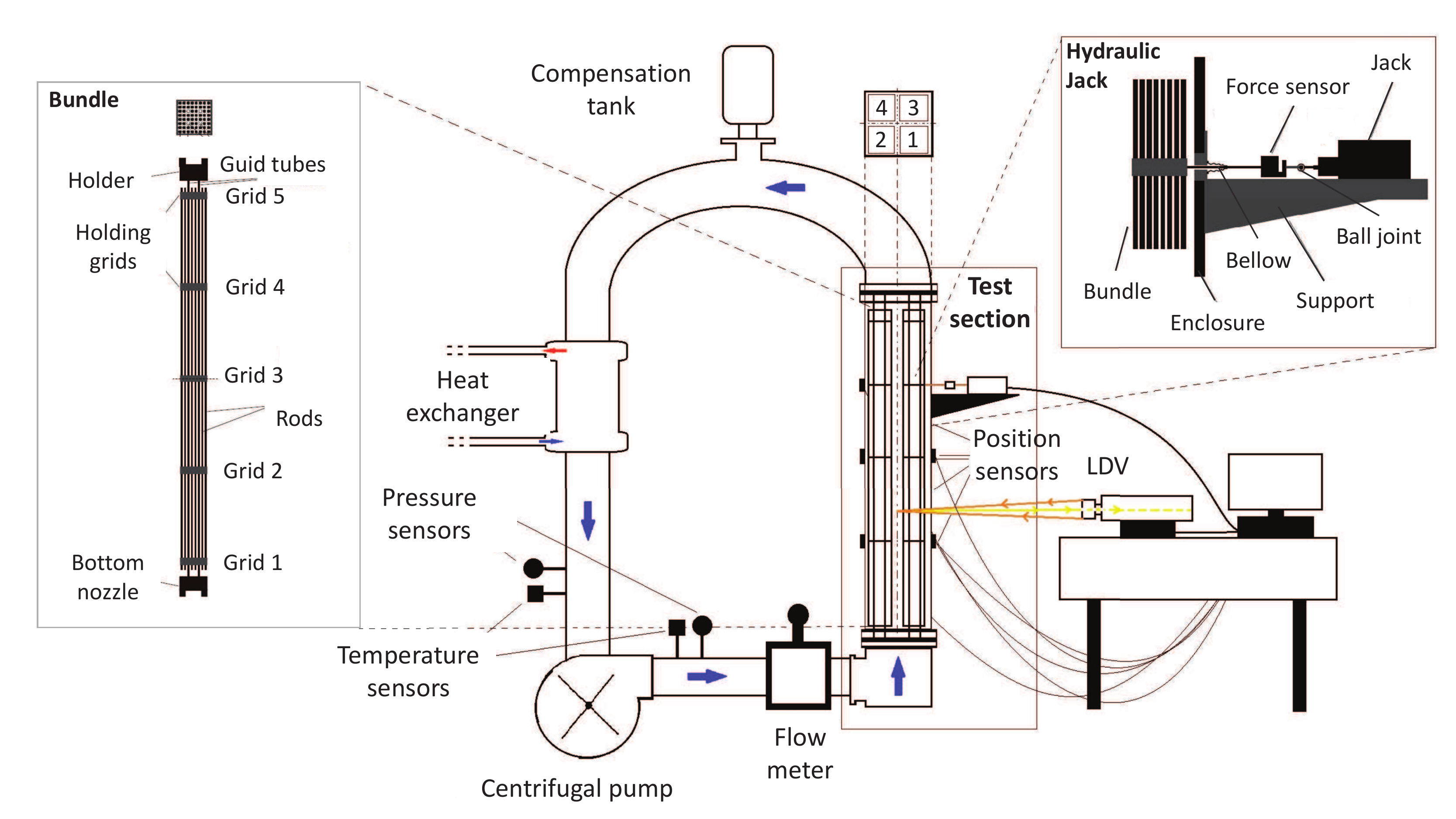}
\caption{Scheme of the ICARE experimental facility.}
\label{fig-icare}
\end{figure}

The ICARE experimental facility is made of a closed water loop powered by a centrifugal pump, the test section, a compensation tank and a heat exchanger (Figure \ref{fig-icare}). The test section is placed in vertical position with a square section of 22.5 cm x 22.5 cm and hosts up to 4 fuel assemblies arranged in a 2 x 2 lattice. The length of the fuel assemblies  is $L=2.57$~m. Each of them is constituted of a squared lattice of $8\times  8$ rods, in which there are 60 rods simulating fuel rods made of acrylic and 4 stainless steel guide tubes. The guide tubes are empty inside, and they are welded to 5 metallic spacer grids along the length of the assembly. The guide tubes have a structural function since they give rigidity to the assembly and they hold together fuel rods. Each assembly has a section of $10.1 \times 10.1\,$cm$^2$, and the 60 rods constituting it have a diameter of $2A=9$~mm with a pin pitch $P/(2A) = 1.33$. The top and bottom of the assembly are rigidly fixed to the guide tubes, and they are fixed to the test section through the lower support plate and the upper support plate.

\begin{figure}
\centering
\includegraphics[width=0.8\textwidth]{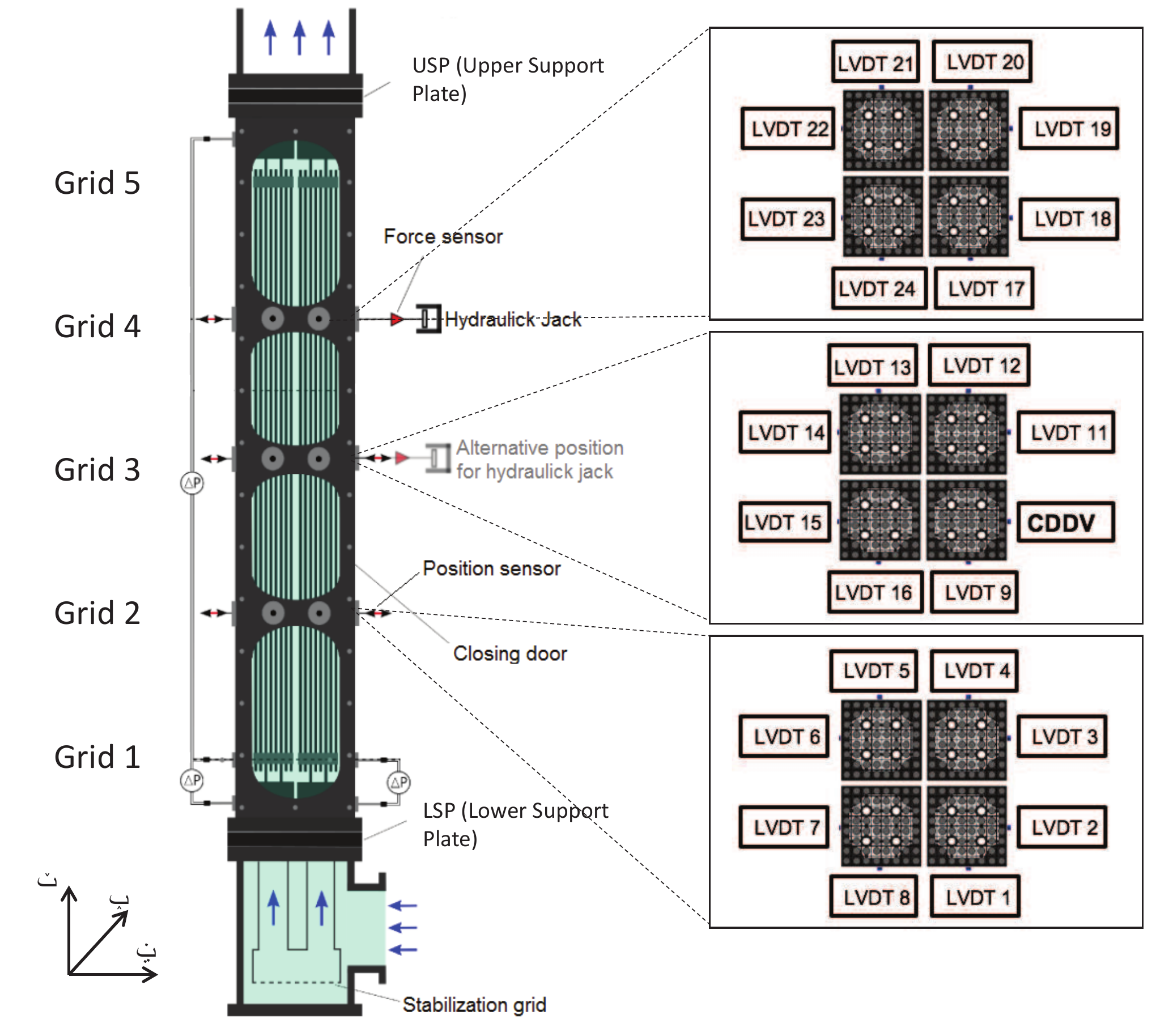}
\caption{Scheme of the displacement sensors on the ICARE test section.}
\label{sens}
\end{figure}

A hydraulic actuator is used to force one of the four fuel assemblies in one direction; it is attached to one of the grids of the fuel assembly through a screw, and a force sensor is installed in between the actuator and the stem. The test section is equipped with 24 linear variable differential transformer (LVDT) position sensors to measure the displacement of each grid in two directions (Figure \ref{sens}). For more details on the ICARE experiment, please refer to \cite{Capanna_ned}.
\subsection{Problem geometry and meshing}
\begin{figure}
\centering
\begin{subfigure}{2.6in}
\centering
\includegraphics[width=2.2in]{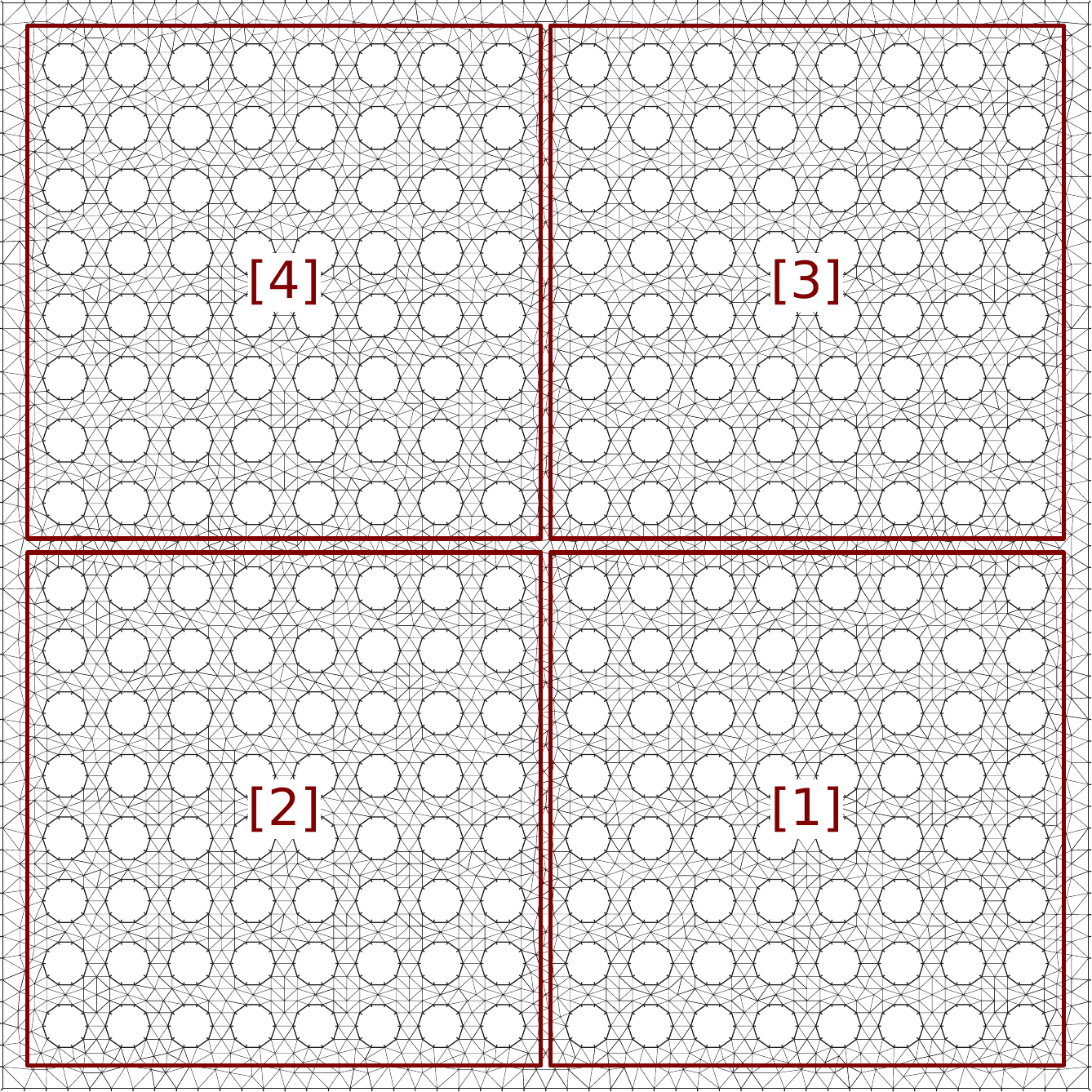}
\caption{4 assemblies in large confinement.}
\label{fig::Meshing_ICARE_4Ass_Large}
\end{subfigure}
\begin{subfigure}{2.6in}
\centering
\includegraphics[width=2.2in]{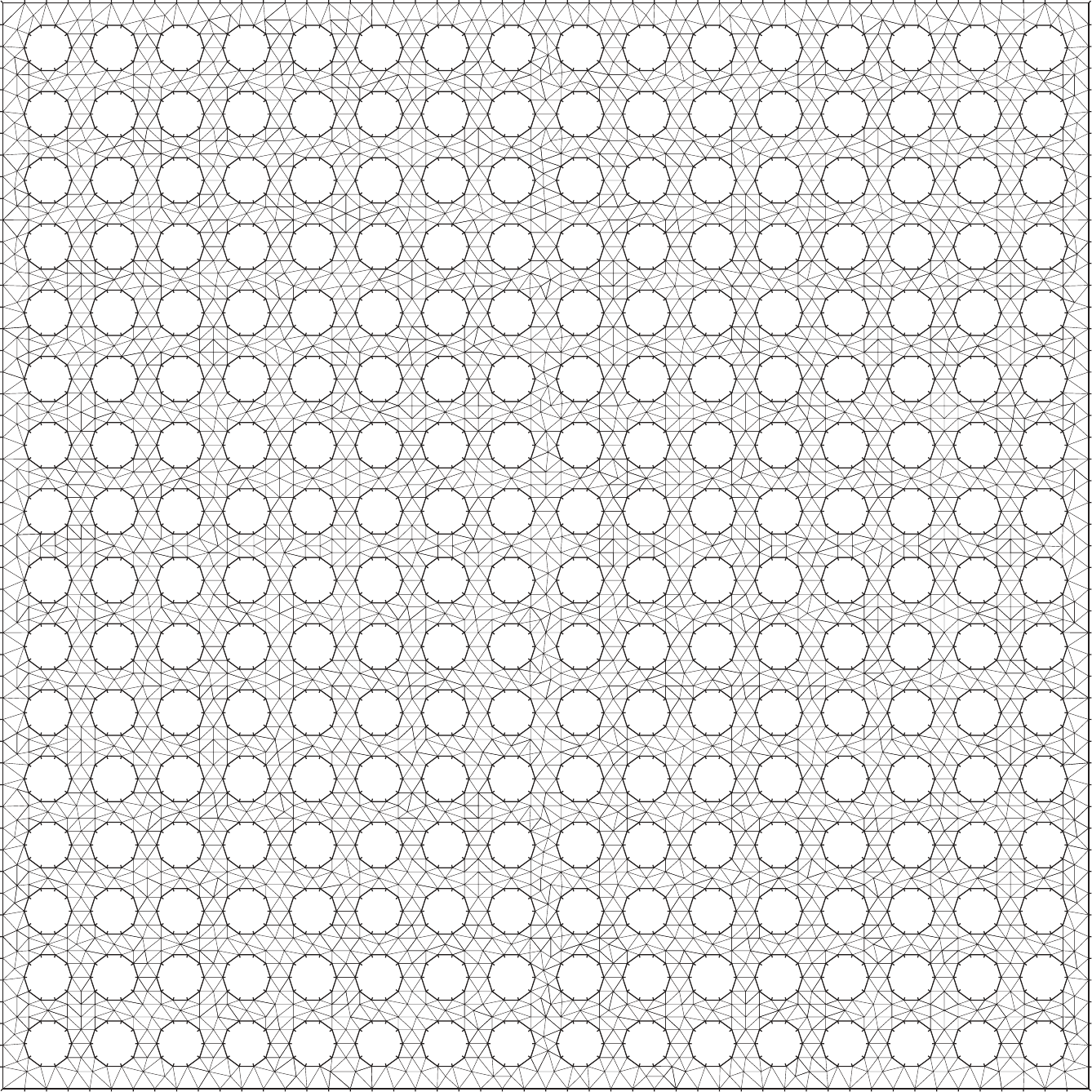}
\caption{4 assemblies in small confinement.}
\label{fig::Meshing_ICARE_4Ass_Small}
\end{subfigure}
\caption{Mesh for the two configurations simulating the ICARE set-up.}
\label{fig::Meshing_ICARE}
\end{figure}

The ICARE geometry is  implemented in the code to allow for comparison with experiments with small (4 mm) or large (8 mm) confinements (see Figures \ref{fig::Meshing_ICARE_4Ass_Large}, \ref{fig::Meshing_ICARE_4Ass_Small}). Confinement size denotes the smallest distance between rods border and casing walls as well as the smallest distance between rods from different assemblies.

Assemblies  are numbered  as depicted in Figure~\ref{fig::Meshing_ICARE_4Ass_Large}.
As in the experiments, the rods of assembly [1] are forced along the $x$-direction and they all exhibit the same displacement $\wx[1](z,t)$. Those rods are assumed not to move along $y$ direction and  the rods in the other assemblies are assumed not to move.

Based on \S\ref{sec::SingleCylinder}, the Helmholtz problem to solve is
\begin{subequations}\label{eq.Helmholtz.assemblies}
\begin{align}
    \Delta\hat{p} &= k^2\hat{p},&\text{in the domain},\\
    \left.\frac{\partial\hat{p}}{\partial n}\right|_{\partial\text{Cas.}} &= 0,
        &\text{on the  walls},\\
    \left.\frac{\partial\hat{p}}{\partial n}\right|_{\dRod[1]} &= \hvx[1](k,t)\cos(\theta),
        &\text{for assembly [1]},\\
    \left.\frac{\partial\hat{p}}{\partial n}\right|_{\dRod[n]} &= 0,
        &\text{for other assemblies},
\end{align}
\end{subequations}
where the superscripts $[n]$ refer to the  assembly number, $\dRod[n]$ denotes the boundaries of rods of assembly $[n]$ (with $\theta$ the local polar angle for each rod) and 
\begin{equation}
    \hvx[1]=\FT{(\partial_t+\partial_z)^2\wx[1]}.
\end{equation}

As in \S\ref{sec::SingleCylinder}, the solution $\hat p$ of the Helmholtz problem \eqref{eq.Helmholtz.assemblies} is proportional to $\hvx[1]$ and one can define added masses $\hat{\mu}_x^{[n]}$ and $\hat{\mu}_y^{[n]}$ from the forces exerted on each assembly
\begin{subequations}
\begin{align}
\hat{f}_x^{[n]}(k,t) 
    & = -\int_{\dRod[n]}\hat p\cos(\theta)\,d\theta 
      = -\hat{\mu}_x^{[n]} \hvx[1],\\
\hat{f}_y^{[n]}(k,t) 
    & = -\int_{\dRod[n]}\hat p\sin(\theta)\,d\theta
      = -\hat{\mu}_y^{[n]} \hvx[1].    
\end{align}
\end{subequations}

In practice, $\hat{\mu}_x^{[n]}$ and $\hat{\mu}_y^{[n]}$ can be calculated by assuming $\hvx[1]=1$ and numerically solving the Helmholtz problem \eqref{eq.Helmholtz.assemblies} for $\hat p$.
The convergence of the calculations depends on the mesh size. We thus conducted a convergence study and found that convergence is ensured when the walls are divided  at least into $150$ sections, and the cylinder borders  into $50$ sections. The calculations presented here have been performed with $200$ divisions on the external walls and $75$ divisions on the cylinder borders, which corresponds to approximately $500\,000$ cells in the domain and a computing time of about $60\,$s.

\begin{figure}
\centering
\includegraphics[width=\textwidth]{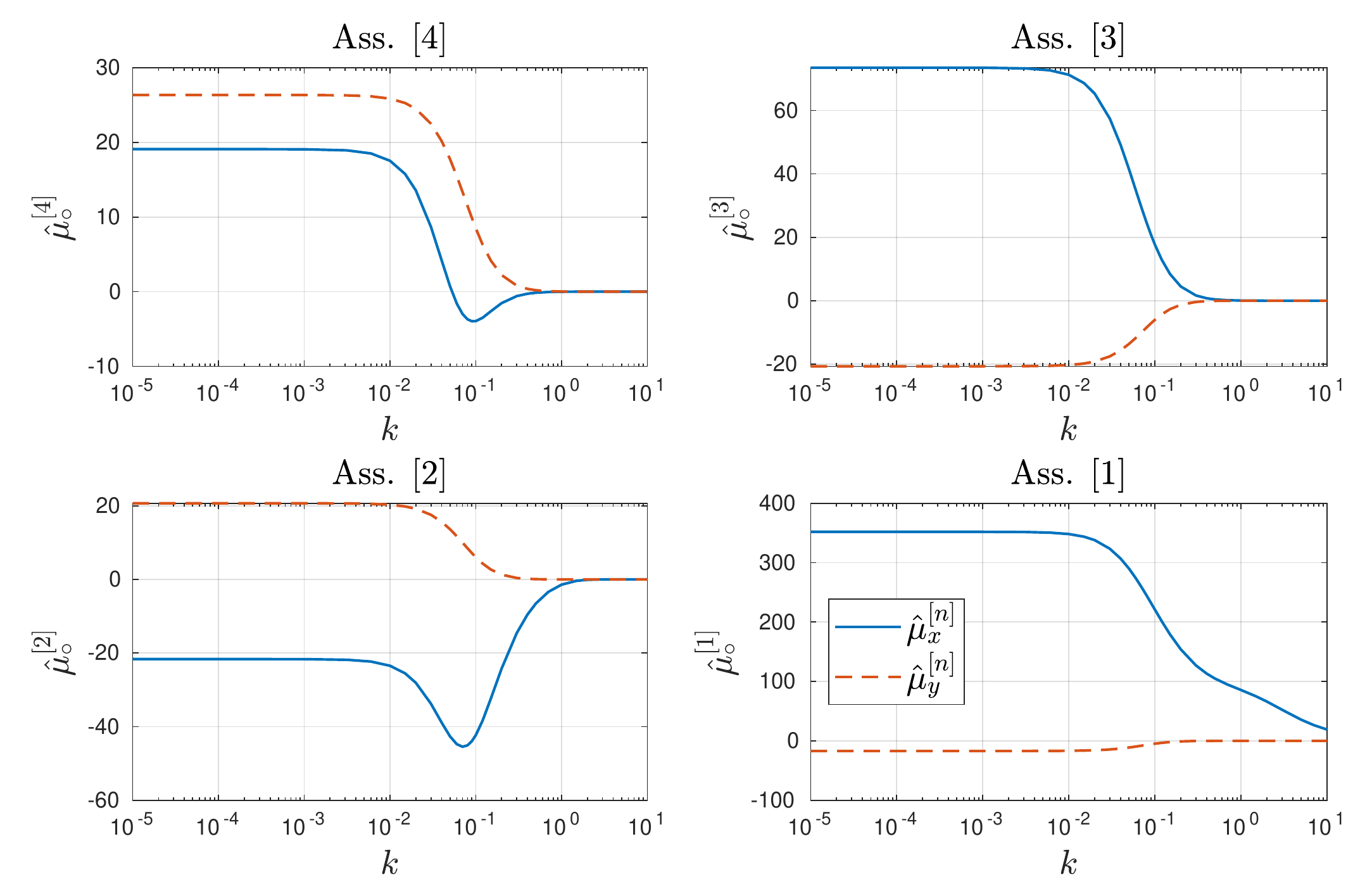}
\caption{Added masses $\hat{\mu}_x^{[n]}$ (blue solid line) and $\hat{\mu}_y^{[n]}$ (red dashed line) on each assembly (in Fourier space) for different values of $k$.}
\label{fig:4x4_8mm_perAssembly_rhatf}
\end{figure}

Figure~\ref{fig:4x4_8mm_perAssembly_rhatf} shows the results of these computations for $k\in [10^{-5}, 10]$ (In Appendix~\ref{sec.qualitative}, we detail these results for individual rods).
As in the single-cylinder case, the added masses do not change much for extreme values of $k$, i.e. $k\lesssim10^{-2}$ or $k\gtrsim1$, except for assembly $[1]$, the forced assembly. For large $k$, the added masses converge to zero.

\subsection{Equations of motion in real space}\label{sec.eqs_motion_real_space}
The functions  $\hat{\mu}_x^{[n]}(k)$ and $\hat{\mu}_y^{[n]}(k)$ can be interpolated using piecewise 6th-order polynomials, which can then be used to calculate ${\mu}_x^{[n]}(z)$ and ${\mu}_y^{[n]}(z)$ with an inverse Fourier transform (appendix~\ref{app:tokeep}).
By analogy with the calculations for a single cylinder done in \S\ref{sec.generalised_coordinates}, we can define added mass, added damping, and added stiffness matrices for the assemblies too. 
There are 24 such matrices: 3 kinds (mass $\mathsfbi{M}$, damping $\mathsfbi{C}$, stiffness $\mathsfbi{K}$), 2 axes ($x$ and $y$, noted $\circ$) and 4 assemblies noted with the superscript $[n]$. 
Similarly to (\ref{eq.added_matrices}\textit{a}--\textit{c}), these matrices are expressed from the added masses ${\mu}_\circ^{[n]}(z)$ 
\begin{subequations}\label{eq.added_matrices_assemblies}
\begin{eqnarray}
    \left(\mathsfbi{M}_\circ^{[n]} \right)_{ij} & = &
    \langle \chi_i, \FTinv{\hat\mu_\circ^{[n]} \hat \chi_j}\rangle =
    \frac{1}{2 \pi}\langle \chi_i, \mu_\circ^{[n]} \star \chi_j\rangle,\\
    \left(\mathsfbi{C}_\circ^{[n]} \right)_{ij} & = &
    \langle \chi_i, \FTinv{2ik\hat\mu_\circ^{[n]} \hat \chi_j}\rangle =
    \frac{1}{2 \pi}\langle \chi_i, 2\mu_\circ^{[n]} \star \chi'_j\rangle,\\
    \left(\mathsfbi{K}_\circ^{[n]} \right)_{ij} & = &
    \langle \chi_i, \FTinv{-k^2\hat\mu_\circ^{[n]} \hat \chi_j}\rangle =
    \frac{1}{2 \pi}\langle \chi_i, \mu_\circ^{[n]} \star \chi''_j\rangle,
\end{eqnarray}
\end{subequations}
where $\chi_i(z)$ is the $i$-th beam eigenmode defined in \eqref{eq.eigenmodes}.

Because it is forced at mid-height, we can assume that the assembly [1] moves mainly on its first beam eigenmode, i.e. $w_x^{[1]}(z,t) = q_x^{[1]}(t) \chi_1(z)$. 
We can further assume that the added mass and added stiffness matrices are diagonal and that the added damping is negligible  (appendix~\ref{app:tokeep}). 
With these hypotheses,  the equation of motion for the $x$-component of the assembly [1] is
\begin{equation}\label{eq.assembly1}
    \left(m_s + m_x^{[1]}\right)\ddot q_x^{[1]} +
    c_v\dot q_x^{[1]} +
    \left(k_s + k_x^{[1]}\right) q_x^{[1]} =
    f_e e^{i\omega t},
\end{equation}
where $m_s=64 m$ is the dimensionless mass of one assembly per unit length, $k_s = 64 b k_1^4$ its stiffness, $c_v$ account for the damping induced by the viscosity effect, $m_x^{[1]}$ and $k_x^{[1]}$ note the component $(1,1)$ of the added matrices $\mathsfbi{M}_x^{[1]}$ and $\mathsfbi{K}_x^{[1]}$
and $f_e e^{i\omega t}$ is the external forcing projected onto the first eigenmode with $f_e$ a scalar. 
$m_x^{[1]}$ and $k_x^{[1]}$ coefficients represent added mass and stiffness encountered by an assembly which moves on its first mode in either $x$ or $y$ direction.
Here, we consider only a weak coupling between the assemblies, such that we can neglect the hydrodynamic forces due to the motion of the other assemblies on assembly $[1]$ (the validity of this assumption will be assessed below).
The solution of \eqref{eq.assembly1} is simply
\begin{equation}
    q_x^{[1]} = h_x^{[1]} f_e e^{i\omega t},
\end{equation}
with the transfer function
\begin{equation}
    h_x^{[1]}(\omega) =\frac{1}
    {-\omega^2 \left(m_s + m_x^{[1]}\right) + i\omega c_v + k_s + k_x^{[1]}}.
\end{equation}

It is possible to estimate the motion of assemblies [2] to [4] by considering the superimposition of the two following cases: still rods undergoing fluids forces induced by assembly [1] on the one hand and moving rods undergoing fluid forces generated by their own movement while other assemblies stand still:
\begin{equation}\label{eq.assemblyn}
    \left(m_s + m_x^{[1]}\right)\ddot q_\circ^{[n]} +
    c_v\dot q_\circ^{[n]} +
    \left(k_s + k_x^{[1]}\right) q_\circ^{[n]} =
    -\left(
       -\omega^2 m_\circ^{[n]} +   k_\circ^{[n]}
    \right) q_x^{[1]} ,
\end{equation}
where the coefficients $m_x^{[1]}$, $c_v$, and $k_x^{[1]}$ appear on the left-hand side because, by symmetry, they also correspond to the pressure force induced by the motion $q_\circ^{[n]}$ on the assembly $[n]$ itself. The right-hand side represents action of assembly [1] motion via fluid.

From (\ref{eq.added_matrices_assemblies}\textit{c}), we see that the added stiffness is $k_\circ^{[n]} = -k_1^2 m_\circ^{[n]}$, with $k_1 = \pi/l$ (since $\chi_i'' = -k_i^2 \chi_i$). 
We will now assume that $\omega \gg k_1$, which is true here because typical values of $\omega$ are around 0.3 and $l \approx 571$.
This means that the added stiffness term $k_\circ^{[n]}$ can be neglected compared to the added mass term $-\omega^2 m_\circ^{[n]}$ in \eqref{eq.assembly1} and \eqref{eq.assemblyn}. The values of the added masses $m_\circ^{[n]}$ are given in table~\ref{matr_added_mass}.
From this table, we see that the ratio of added masses, $m_\circ^{[n]} / m_x^{[1]}$ is at most 21\% (for $n>1$). This justifies to neglect the hydrodynamic forces due to the motion of the assemblies $[n]$ onto assembly $[1]$, which can be estimated as the square of this ratio and thus of order 4\% at most.  

\begin{table}
\begin{center}
\caption{Added mass matrix coefficients for small and large confinement} \label{matr_added_mass}
\begin{tabular}{c cccc cccc}
 confinement & 
    $m_x^{[1]}$ & $m_x^{[2]}$ & $m_x^{[3]}$ & $m_x^{[4]}$ & 
    $m_y^{[1]}$ & $m_y^{[2]}$ & $m_y^{[3]}$ & $m_y^{[4]}$\\
\hline 
large & 350 & -22.1 & 72.7 & 18.6 & -17.0 & 20.5 & -20.5 & 26.2\\
small & 384 & -24.0 & 79.0 & 21.3 & -22.4 & 25.3 & -25.3 & 29.3
\end{tabular}
\end{center}
\end{table}

Under these hypotheses, we can calculate the displacement ratio at the forcing frequency $\omega$
\begin{equation}
    \frac{q_\circ^{[n]}}{q_x^{[1]}} \approx
    -\frac{-\omega^2 m_\circ^{[n]}}{-\omega^2 \left(m_s + m_x^{[1]}\right) + i\omega c_v + k_s},
\end{equation}
which become at the resonant frequency $\omega_\mathrm{res}$
\begin{equation}\label{rap_amp_th}
    \frac{q_\circ^{[n]}}{q_x^{[1]}} \approx -i \omega_\mathrm{res} \frac{m_\circ^{[n]}}{c_v}, \quad \mbox{with }
    -\omega_\mathrm{res}^2 \left(m_s + m_x^{[1]}\right) + k_s = 0, \quad
\end{equation}
note that the latter equation is applicable only for $n\neq$1.

%
\section{Comparison with experimental results}\label{sec:Experiments}
%
This section compares results obtained using the simplified model and results from an experimental campaign on ICARE facility described in \S\ref{sse::ICARE}.
Experiments are performed imposing a sine sweep ranging from 0 to 10 Hz with a constant amplitude of 1 mm and an axial fluid velocity of 1 m/s when water is present. To catch the transfer function of a system using a swept sinus excitation, the sweep velocity needs to be small enough to avoid any transient phenomena and contamination of different harmonics in the system. For these experiment a sweep velocity of 0.05 Hz is applied, respecting the international standard ISO-7626 indications. The frequency range embraces assembly [1] first and second mode in water.

\subsection{Added mass}
Let us consider the transfer function $H^{[1]}_{exp}$ between the displacement $W^{[1]}_{x}$ along $x$ of the grid with vertical position $Z_w$, and the force $F_{g}$ applied by the actuator on grid with position $Z_f$. Since the data are acquired with a swept sinus excitation, the transfer functions are calculated using the cross correlation product, and are defined as 

\begin{eqnarray}
	H^{[1]}_{exp}=\left\|\frac{W^{[1]}_{x} \star W^{[1]}_{x}}{W^{[1]}_{x} \star F_{x}}\right\|
	\label{Transfer_f}
\end{eqnarray}

\noindent 
where $W^{[1]}_{x} \star W^{1]}_{x}$ is the spectral autocorrelation of the displacement signal of the $1$-{st} assembly along the $x$ direction and $W^{[1]}_{x} \star F_{x}$ is the spectral cross correlation between the displacement signal of the $1$-st assembly and the force imposed on the first assembly. Cross correlation functions are calculated using an Hamming windowing filter (5000 points per windows) to clean up the noise from the signal.

The theoretical transfer function modulus for a one degree of freedom system is given by :
\begin{eqnarray}
	H^{[1]}_{theo}(\Omega)=\frac{1}{\sqrt{(K_e-\Omega^2M_e)^2+\Omega^2C_e^2}},
\end{eqnarray}
where $M_e$, $C_e$ and $K_e$ are the coefficients of mass, damping and stiffness identified and $\Omega$ is the circular frequency.

Assembly [1] is excited around its first mode; to identify its dynamical properties, one has to find $M_e$, $C_e$ and $K_e$ that make $H^{[1]}_{theo}$ fit the experimental data $H^{[1]}_{exp}$. This can be easily achieved by a conjugate gradient optimization method.

Using an Euler-Bernoulli beam model for the rods as described in Eq.~\eqref{Eq::Euler_Bernoulli}, one can write the virtual work of assembly [1]:

\begin{equation}
\begin{split}
M_t\int^{L}_{0}\partial^2_T W^{[1]}_{x}\delta WdZ+C_t\int^{L}_{0}\partial_T W^{[1]}_{x}\delta WdZ+B_t\int^{L}_{0}\partial^2_Z W^{[1]}_{x}\partial^2_Z\delta WdZ\qquad\\%
\qquad=\int_0^LF_{g}\delta_{Z_f}\delta W\,dZ,
\end{split}
\end{equation}

with
\begin{equation}
M_t=M_s+M_f, \quad B_t=B+B_f,
\end{equation}
where $M_f$ and $B_f$ are respectively the participation of the fluid into mass and stiffness, $C_t$ is a linear damping coefficient and the sum accounts for the 64 rods inside the assembly.

When the structure vibrates on its $j$-th natural mode 
\begin{equation}
W^{[1]}_{x}(Z,T)=N_j(T) X_j(Z),
\end{equation}
with $X_j(Z)=\chij(Z/A)$, one can rewrite the energy of the system :
\begin{equation}
M_t\int^{L}_{0}X_j^2dZ\,\partial^2_T N_j+C_t\int^{L}_{0}X_j^2dZ\,\partial_T N_j+B_t\int^{L}_{0}(\partial^2_Z X_j)^2dZ\,N_j=F_{g}X_j(Z_f).\label{ident_exp}
\end{equation}

Noting that the experimental signal is 
\begin{equation}
W^{[1]}_{g}=N_1(T) X_1(Z_w),
\end{equation}
one can relate the  mass per unit of length to the identified coefficient:
\begin{equation}
M_t=\frac{X_1(Z_f)X_1(Z_w)}{\int^{L}_{0}X_1^2dZ}M_e.
\end{equation}

Measurements take into account the effect of the fluid added mass $M_f$ and the mass of the structure $M_s$, therefore to isolate the fluid added mass the participation of the structure is subtracted based on experimental results obtained without water. Finally, one then obtains the results listed in Table~\ref{res_m_added} where added mass are provided for the whole assembly:
\begin{equation}
    M^{[1]}_x = M_f.
\end{equation}
To get a comparison with results from numerical simulation obtained in \S\ref{sss::MultipleCylinders_Assembly1}, the following transformation is applied:
\begin{equation}
    M^{[1]}_x = \rho A^2 \, m_x^{[1]},
\end{equation}
with $\rho=1000\,$kg$\,$m$^{-3}$ the water density.

Table \ref{tab_res} shows the values of added mass for large and narrow confinement given by experiments and simulations. Both agree on the increase of added mass when the confinement gets narrower. Experiments gives a large increase, more than double when simulation predict an increase of about 10\%. Although the order of magnitude is respected the differences observed could question the model. To properly analyse these results, one has to keep in mind that the methodology used to obtained the experimental results is subjected to uncertainties coming from the accuracy of measurement, the optimization process and the fact that we try to identify constants of a linear one degree-of-freedom system from a complex nonlinear system. As a result one could estimate that the added mass coefficient given by the simulations is a reasonable estimation.

\begin{table}
\begin{center}
\caption{Values of added mass $M^{[1]}_x$}\label{res_m_added}\label{tab_res}
\begin{tabularx}{130mm}{>{\centering}X>{\centering}p{4cm}>{\centering}X}
      \rule[-1mm]{0mm}{5mm} $M^{[1]}_x$ & \textsc{experiment}  & \textsc{simulation} \tabularnewline
      \hline
      \rule[-3mm]{0mm}{10mm}Large confinement & 4.06 kg/m  & 7.09 kg/m \tabularnewline 
      \rule[-3mm]{0mm}{10mm}Narrow confinement & 10.36 kg/m & 7.78 kg/m \tabularnewline 
\end{tabularx}
\end{center}
\end{table}

No experimental identification can be made on added stiffness as it is very low compared to structure initial stiffness. To give orders of magnitude, the measured stiffness of a fuel assembly $B_t=2300$~N.m$^2$ while considering the estimation of added stiffness from numerical simulation in the large confinement case obtained in \S\ref{sss::MultipleCylinders_Assembly1}, $k^{[1]}_x \rho U^2A^4/(k_1^4)=4.7$~N.m$^2$. Experimental results are not accurate enough to measure this less than 0.5\% added stiffness effect.


\subsection{Coupling}
To characterise the coupling between the fuel assemblies induced by the fluid, let us consider the transfer function between the imposed displacement and the displacement of the other fuel assemblies at the third grid level in the two transverse directions. Similarly to \eqref{Transfer_f} transfer functions are calculated as 

\begin{eqnarray}
	H^{[n]}_{\circ}=\left\|\frac{W^{[n]}_{\circ} \star W^{[n]}_{\circ}}{W^{[n]}_{\circ} \star W^{[1]}_{x}}\right\|,\ \cinxy,
\end{eqnarray}

\noindent 
with $W^{[n]}$ being the displacement signal of the $n$-{th} assembly along the $\circ$ direction.

For each transfer function the maximum value, which occurs at the first natural frequency is accounted for. These experimental results are compared to displacement ratio given by \eqref{rap_amp_th}, with the addedd mass coefficients $m_\circ^{[n]}$ given in table \ref{matr_added_mass} and $c_v$ and $\omega_\mathrm{res}$  estimated from identification of dynamical parameters using \eqref{ident_exp} as follows
\begin{equation}
C_v=\frac{X_1(Z_f)X_1(Z_w)}{\int^{L}_{0}X_1^2dZ}C_e,  \quad
c_v=\frac{C_v}{\rho UA},  \quad
\omega_\mathrm{res}=\frac{A}{U}\sqrt{\frac{K_e}{M_e}}.
\end{equation}
Experimental values give $c_v=69.2$ and $\omega_\mathrm{res}=0.12$.

\begin{figure} 
\centering
\includegraphics[width=\textwidth]{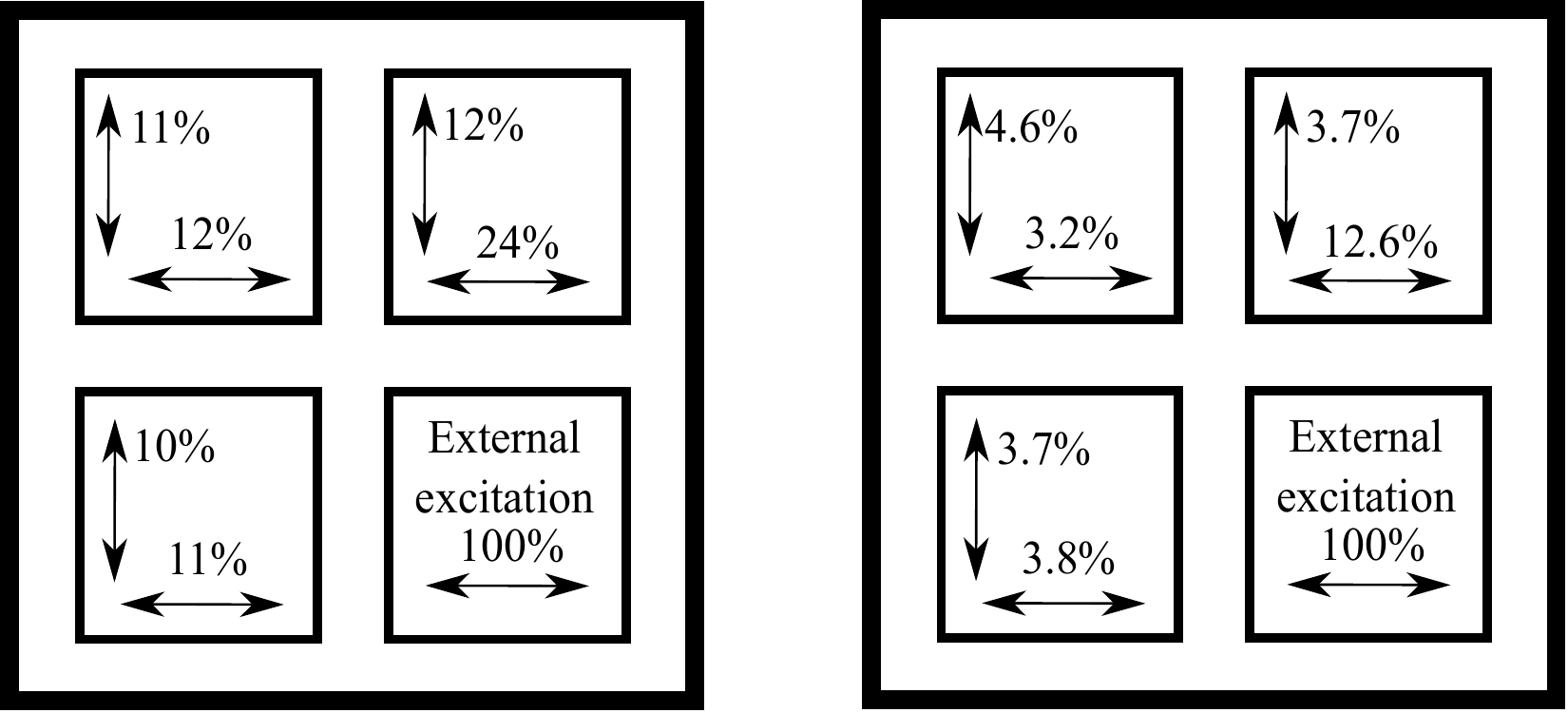}
\caption{Displacement ratios obtained by experimental data (left) and approximated from numerical simulations (right) for 4 fuel assemblies in large confinement at first natural frequency.}
\label{fig:CouplingRatios}
\end{figure}

Figure~\ref{fig:CouplingRatios} compare graphically the  displacement ratios as observed in the experiments with the ratios predicted by \eqref{rap_amp_th} with these values of $c_v$ and $\omega_\mathrm{res}$. Experiments show a ratio of about 10 \% for most of the fuel assemblies in the two directions except for the fuel assembly [3] which shows a coupling about 20 \% in the direction of forcing. Although simulations seems to underestimate the values of coupling, they give a good order of magnitude. Moreover, the general pattern is well reproduced with a significantly more important coupling for fuel assembly [3] in the direction of forcing.
Therefore, the methodology developed seems to be appropriate to simulate the coupling between fuel assemblies.

%
\section{Conclusions}\label{sec:Conclusion}
In this paper a new model for fluid-structure interaction of PWR fuel assemblies is proposed. The model used the well known potential flow theory and the equation are solved by using a Fourier approach. This approach leads to the important result of relating the fluid-structure interaction forces directly to the displacement of the structure itself. This result is an important achievement since it allows to drastically reduce computational time, avoiding to solve fluid equations. As a drawback it should be considered that the assumption of potential flow leads to the complete negligence of the fuel grids, which introduce vorticity in the flow. 

Calculations with potential flow model are first performed for single cylinder geometry and results have been compared with reference works in literature. The model showed perfect agreement with the Slender Body Model and it also shows good predictions respect to the confinement size. This validation has proven the model, and thus the mathematical approach used to solve the equations, to be consistent and reliable. The model was thus improved for multiple cylinder geometries. 

Finally, in order to validate the multiple cylinder model, calculations simulating the ICARE geometry are discussed and compared with experimental results.  Modal parameters are identified as functions of the flow rates and compared with experimental ones. The model reasonably  fits experimental data for the added mass and gives a good estimation of the coupling between fuel assemblies.



Many perspectives are opened as a result of this work. As discussed in this paper the model which has been implemented does not account for viscous forces. An improvement that would be of fundamental importance is to introduce with empirical factors the viscous forces in the model. One possible way in order to introduce viscosity in the model is to consider the axial water flow as the sum of two flows: a potential bulk flow and a viscous flow which has a defined empirical distribution. In this way the viscous forces are taken into account with an empirical formulation, and it would be extremely interesting to assess the effects of such improvements, especially on the added damping coefficient estimation.


\section*{Acknowledgement}
The authors are grateful for the financial support of the Electricit\'e de France (EDF) and FRAMATOME.


\section*{Declaration of Interests}
The authors report no conflict of interest.

%
\bibliographystyle{jfm}
\bibliography{Main_JFM_Potential_Flow}

\begin{thebibliography}{41}
\expandafter\ifx\csname natexlab\endcsname\relax\def\natexlab#1{#1}\fi
\def\au#1{#1} \def\ed#1{#1} \def\yr#1{#1}\def\at#1{#1}\def\jt#1{\textit{#1}}
  \def\bt#1{#1}\def\bvol#1{\textbf{#1}} \def\vol#1{#1} \def\pg#1{#1}
  \def\publ#1{#1}\def\arxiv#1{#1}\def\org#1{#1}\def\st#1{\textit{#1}}

\bibitem[Bauchau \& Craig(2009)]{Bauchau}
{\sc \au{Bauchau, O.~A.} \& \au{Craig, J.~I.}} \yr{2009} {\em Structural
  {A}nalysis: {W}ith {A}pplications to {A}erospace {S}tructures\/}.
  \publ{Springer}.

\bibitem[Blevins(2015)]{Blevins}
{\sc \au{Blevins, R.D.}} \yr{2015} {\em Formulas for Dynamics, Acoustics and
  Vibration\/}.  \publ{John Wiley \& Sons, Ltd}.

\bibitem[Brennen(1982)]{Brennen_82}
{\sc \au{Brennen, C.E.}} \yr{1982}  \at{A review of added mass and fluid
  inertial forces}.  \jt{Technical Report Naval Civil Engineering Laboratory} .

\bibitem[Capanna(2018)]{Capanna_PhD}
{\sc \au{Capanna, R.}} \yr{2018}  \at{Modelling of fluid structure interaction
  by potential flow theory in a {PWR} under seismic excitation}.  \jt{PhD
  Thesis} .

\bibitem[Capanna {\em et~al.\/}(2019)Capanna, Ricciardi, Eloy \&
  Sarrouy]{Capanna_ned}
{\sc \au{Capanna, R.}, \au{Ricciardi, G.}, \au{Eloy, C.} \& \au{Sarrouy, E.}}
  \yr{2019}  \at{Experimental study of fluid structure interaction on fuel
  assemblies on the icare experimental facility}.  \jt{Nuclear Engineering and
  Design}  \bvol{352}.

\bibitem[Chen(1985)]{Chen_85}
{\sc \au{Chen, S.S.}} \yr{1985}  \at{Flow-induced vibration of circular
  cylindrical structures}.  \jt{Argonne National Lab}  \bvol{ANL 85-81}.

\bibitem[Chen \& Wamnsganss(1972)]{Chen_72}
{\sc \au{Chen, S.S.} \& \au{Wamnsganss, M.W.}} \yr{1972}
  \at{Parallel-flow-induced vibration of fuel rods}.  \jt{Nuclear Engineering
  and Design}  \bvol{18}~(2),  \pg{253--278}.

\bibitem[Clasen(1972)]{Clasen_72}
{\sc \au{Clasen, P.}} \yr{1972}  \at{Die 'hydrodynamische masse' eines rohrs in
  einem rohrbundel}.  \jt{Forshung im IngenieurwesenNuclear Engineering and
  Design}  \bvol{38}~(2),  \pg{33--37}.

\bibitem[De~Langre {\em et~al.\/}(2007)De~Langre, Paa\"idoussis, Doar\'e \&
  Modarres~Sadeghi]{De_Langre}
{\sc \au{De~Langre, E.}, \au{Paa\"idoussis, M.P.}, \au{Doar\'e, O.} \&
  \au{Modarres~Sadeghi, Y.}} \yr{2007}  \at{Flutter of long flexible cylinders
  in axial flow}.  \jt{Journal of Fluid Mechanics}  \bvol{571},  \pg{371--389}.

\bibitem[De~Ridder {\em et~al.\/}(2013)De~Ridder, Degroote, Van~Tichelen \&
  Vierendeels]{De_Ridder_13}
{\sc \au{De~Ridder, J.}, \au{Degroote, J.}, \au{Van~Tichelen, K.} \&
  \au{Vierendeels, J.}} \yr{2013}  \at{Modal characteristics of a flexible
  cylinder in turbulent axial flow from numerical simulations}.  \jt{Journal of
  Fluids and Structures}  \bvol{43},  \pg{110--123}.

\bibitem[De~Ridder {\em et~al.\/}(2017)De~Ridder, Degroote, Van~Tichelen \&
  Vierendeels]{De_Ridder_17}
{\sc \au{De~Ridder, J.}, \au{Degroote, J.}, \au{Van~Tichelen, K.} \&
  \au{Vierendeels, J.}} \yr{2017}  \at{Predicting modal characteristics of a
  cluster of cylinders in axial flow: from potential flow solutions to coupled
  {CFD-CSM} calculations}.  \jt{Journal of Fluids and Structures}  \bvol{74},
  \pg{90--110}.

\bibitem[De~Ridder {\em et~al.\/}(2015)De~Ridder, Doar\'e, Degroote,
  Van~Tichelen, Schuurmans \& Vierendeels]{De_Ridder_15}
{\sc \au{De~Ridder, J.}, \au{Doar\'e, O.}, \au{Degroote, J.}, \au{Van~Tichelen,
  K.}, \au{Schuurmans, J.} \& \au{Vierendeels, J.}} \yr{2015}  \at{Simulating
  the fluid forces and fluid-elastic instabilities of a clamped-clamped
  cylinder in a turbulent flow}.  \jt{Journal of Fluids and Structures}
  \bvol{55},  \pg{139--154}.

\bibitem[Doar\'e {\em et~al.\/}(2011{\natexlab{{\em a\/}}})Doar\'e, Mano \&
  Bilbao~Ludena]{Doare_2}
{\sc \au{Doar\'e, O.}, \au{Mano, D.} \& \au{Bilbao~Ludena, J.~C.}}
  \yr{2011{\natexlab{{\em a\/}}}}  \at{Effect of spanwise confinement on flag
  flutter: experimental measurements}.  \jt{Physics of Fluids, American
  Institute of Physics}  \bvol{23},  \pg{11704}.

\bibitem[Doar\'e {\em et~al.\/}(2011{\natexlab{{\em b\/}}})Doar\'e, Sauzade \&
  Eloy]{Doare_2011}
{\sc \au{Doar\'e, O.}, \au{Sauzade, M.} \& \au{Eloy, C.}}
  \yr{2011{\natexlab{{\em b\/}}}}  \at{Flutter of an elastic plate in a channel
  flow: confinement and finite-size effects}.  \jt{Journal of Fluids and
  Structures}  \bvol{27}~(1),  \pg{76--88}.

\bibitem[Du~Buat(1786)]{DuBuat}
{\sc \au{Du~Buat, P.L.G.}} \yr{1786}  \at{Principles d'hydraulique}.
  \jt{Imprimerie du Monsieur} .

\bibitem[Eloy {\em et~al.\/}(2010)Eloy, Doar\'e, Duchemin \& L.]{Doare_3}
{\sc \au{Eloy, C.}, \au{Doar\'e, O.}, \au{Duchemin, L.} \& \au{L.,
  Schouveiler}} \yr{2010}  \at{A unified introduction to fluid mechanics of
  flying and swimming at high reynolds number}.  \jt{Experimental Mechanics}
  \bvol{50}~(9),  \pg{1361--1366}.

\bibitem[Eloy {\em et~al.\/}(2008)Eloy, Lagrange, Souilliez \&
  Schouveiler]{Eloy_2008}
{\sc \au{Eloy, C.}, \au{Lagrange, R.}, \au{Souilliez, C.} \& \au{Schouveiler,
  L.}} \yr{2008}  \at{Aeroelastic instability of cantilevered flexible plates
  in uniform flow}.  \jt{Journal of Fluid Mechanics}  \bvol{611},
  \pg{97--106}.

\bibitem[Eloy {\em et~al.\/}(2007)Eloy, Souillez \& Schouveiler]{Eloy_2007}
{\sc \au{Eloy, C.}, \au{Souillez, C.} \& \au{Schouveiler, L.}} \yr{2007}
  \at{Flutter of a rectangulare plate}.  \jt{Journal of Fluids and Structures}
  \bvol{23}~(6),  \pg{904--919}.

\bibitem[Gosselin \& De~Langre(2011)]{Gosselin_2011}
{\sc \au{Gosselin, F.P.} \& \au{De~Langre, E.}} \yr{2011}  \at{Drag reduction
  by reconfiguration of a poroeslatic system}.  \jt{Journal of Fluids and
  Structures}  \bvol{27}~(7),  \pg{1111--1123}.

\bibitem[Guo \& Pa\"idoussis(2000)]{Guo_2000}
{\sc \au{Guo, C.Q.} \& \au{Pa\"idoussis, M.~P.}} \yr{2000}  \at{Stability of
  rectangular plates with free side-edges in two-dimensional inviscid channel
  flow}.  \jt{Journal of Applied Mechanics}  \bvol{67}~(1),  \pg{171--176}.

\bibitem[Hecht(2012{\natexlab{{\em a\/}}})]{FreeFem_0}
{\sc \au{Hecht, F.}} \yr{2012{\natexlab{{\em a\/}}}}  \at{Freefem++ v.3.19-1}.
  \jt{Universite {P}ierre et {M}arie {C}urie.
  http://www.freefem.org/ff++/ftp/freefem++doc.pdf} .

\bibitem[Hecht(2012{\natexlab{{\em b\/}}})]{FreeFem_1}
{\sc \au{Hecht, F.}} \yr{2012{\natexlab{{\em b\/}}}}  \at{New development in
  {F}ree{F}em++}.  \jt{Journal of Numerical Mathematics}  \bvol{20}~(3-4),
  \pg{251--265}.

\bibitem[Lamb(1895)]{Lamb_1895}
{\sc \au{Lamb, H.}} \yr{1895}  \at{Hydrodynamics}.  \jt{Cambridge University
  Press} .

\bibitem[Lighthill(1960{\natexlab{{\em a\/}}})]{Lighthill_Slender_Body}
{\sc \au{Lighthill, M.J.}} \yr{1960{\natexlab{{\em a\/}}}}  \at{Mathematics and
  {A}eronautics}.  \jt{The Journal of the Royal Aeronautical Society}
  \bvol{64}~(595),  \pg{375--394}.

\bibitem[Lighthill(1960{\natexlab{{\em b\/}}})]{Lighthill_1960}
{\sc \au{Lighthill, M.J.}} \yr{1960{\natexlab{{\em b\/}}}}  \at{Note on the
  swimming of slender fish}.  \jt{Journal of Fluid Mechanics}  \bvol{9}~(2),
  \pg{305--317}.

\bibitem[Michelin \& Smith(2009)]{Michelin_2009}
{\sc \au{Michelin, S.} \& \au{Smith, S.L.}} \yr{2009}  \at{Linear stability
  analysis of coupled parallel flexible plates in an axial flow}.  \jt{Journal
  of Fluids and Structures}  \bvol{25}~(7),  \pg{1136--1157}.

\bibitem[Morison {\em et~al.\/}(1950)Morison, Johnson \& Schaaf]{Morison}
{\sc \au{Morison, J.R.}, \au{Johnson, J.W.} \& \au{Schaaf, S.A.}} \yr{1950}
  \at{The force distribution exerted by surface waves on piles}.  \jt{Journal
  of Petroleum Technology}  \bvol{2}~(5),  \pg{149--154}.

\bibitem[Pa\"idoussis(1966{\natexlab{{\em a\/}}})]{Paid_1966_Theo}
{\sc \au{Pa\"idoussis, M.~P.}} \yr{1966{\natexlab{{\em a\/}}}}  \at{Dynamics of
  slender cylinders in axial flow: {P}art 1 {T}heory}.  \jt{Journal of Fluid
  Mechanics}  \bvol{26}~(4),  \pg{717--736}.

\bibitem[Pa\"idoussis(1966{\natexlab{{\em b\/}}})]{Paid_1966_Exp}
{\sc \au{Pa\"idoussis, M.~P.}} \yr{1966{\natexlab{{\em b\/}}}}  \at{Dynamics of
  slender cylinders in axial flow: {P}art 2 {E}xperiments}.  \jt{Journal of
  Fluid Mechanics}  \bvol{26}~(4),  \pg{737--751}.

\bibitem[Pa\"idoussis(1973)]{Paid_1973}
{\sc \au{Pa\"idoussis, M.~P.}} \yr{1973}  \at{Dynamics of cylindrical
  structures subjected to axial flow}.  \jt{Journal of Sound and Vibration}
  \bvol{29}~(3),  \pg{365--385}.

\bibitem[Pa\"idoussis(1979)]{Paid_1979_Clusters}
{\sc \au{Pa\"idoussis, M.~P.}} \yr{1979}  \at{The dynamics of clusters of
  flexible cylinders in axial flow: theory and experiments}.  \jt{Journal of
  Sound and Vibration}  \bvol{65}~(3),  \pg{391--417}.

\bibitem[Pa\"idoussis \& Pettigrew(1979)]{Paid_1979}
{\sc \au{Pa\"idoussis, M.~P.} \& \au{Pettigrew, M.J.}} \yr{1979}  \at{Dynamics
  of flexible cylinders in axisymetrically confined axial flow}.  \jt{Journal
  of Applied Mechanics}  \bvol{46}~(1),  \pg{37--44}.

\bibitem[Pa\"idoussis \& Suss(1977)]{Paid_1977_b}
{\sc \au{Pa\"idoussis, M.~P.} \& \au{Suss, S.}} \yr{1977}  \at{Stability of a
  cluster of flexible cylinders in bounded axial flow}.  \jt{Journal of Applied
  Mechanics}  \bvol{44}~(3),  \pg{401--408}.

\bibitem[Pa\"idoussis {\em et~al.\/}(1977)Pa\"idoussis, Suss \&
  Pustejovsky]{Paid_1977_a}
{\sc \au{Pa\"idoussis, M.~P.}, \au{Suss, S.} \& \au{Pustejovsky, M.}} \yr{1977}
   \at{Free vibration of clusters of cylinders in liquid-filled channels}.
  \jt{Journal of Sound and Vibration}  \bvol{55}~(3),  \pg{443--459}.

\bibitem[Ricciardi {\em et~al.\/}(2009)Ricciardi, Bellizzi, Collard \&
  B.]{Ricciardi_2009}
{\sc \au{Ricciardi, G.}, \au{Bellizzi, S.}, \au{Collard, B.} \& \au{B.,
  Cochelin}} \yr{2009}  \at{Modelling {P}ressurized {W}ater {R}eactor cores in
  terms of porous media}.  \jt{Journal of Fluids and Structures}
  \bvol{25}~(1),  \pg{112--133}.

\bibitem[Schouveiler \& Eloy(2009)]{Schouveiler_2009}
{\sc \au{Schouveiler, L.} \& \au{Eloy, C.}} \yr{2009}  \at{Coupled flutter of
  parallel plates}.  \jt{Physics of Fluids}  \bvol{21}~(8).

\bibitem[Singh {\em et~al.\/}(2012)Singh, Michelin \& De~Langre]{Singh_2012}
{\sc \au{Singh, K.}, \au{Michelin, S.} \& \au{De~Langre, E.}} \yr{2012}
  \at{Energy harvesting from fluid-elastic instabilities of a cylinder}.
  \jt{Journal of Fluids and Structures}  \bvol{30},  \pg{159--172}.

\bibitem[Taylor(1952)]{Taylor_52}
{\sc \au{Taylor, G.I.}} \yr{1952}  \at{Analysis of the swimming of long and
  narrow animals}.  \jt{Proceedings of the Royal Society}  \bvol{214}~(1117),
  \pg{158--183}.

\bibitem[Triantafyllou \& Cheryssostomidis(1985)]{Triantafyllou_85}
{\sc \au{Triantafyllou, G.I.} \& \au{Cheryssostomidis, C.}} \yr{1985}
  \at{Stability of a string in axial flow}.  \jt{ASME Journal of Energy
  Ressources Technology}  \bvol{107}~(421),  \pg{421--425}.

\bibitem[Wendel(1956)]{Wendel_56}
{\sc \au{Wendel, K.}} \yr{1956}  \at{Hydrodynamic masses and hydrodynamic
  moments of inertia}.  \jt{TMB Translation}  \bvol{260}.

\bibitem[Wu(2001)]{Wu_2001}
{\sc \au{Wu, T.Y.}} \yr{2001}  \at{Mathematical biofluiddynamics and
  mechanophysiology of fish locomotion}.  \jt{Mathematical Methods in the
  Applied Sciences}  \bvol{24}~(17-18),  \pg{1464--1541}.

\end{thebibliography}
%
\appendix

\section{FreeFEM++ code for one cylinder}\label{app:FreeFEM1Cylinder}
\footnotesize
\lstset{numbers=left, numberstyle=\tiny, stepnumber=5, numbersep=5pt,basicstyle=\scriptsize,stringstyle=\ttfamily}
\begin{lstlisting}
real k = 0.01;
real lx = 1.5;
real ly = 2;
real A = 1.;

// Geometry
border Cxb(t=0, 1){x=A+(lx-A)*t^(2); y=0*t; label=1;}
border Cyr(t=0, 1){x=lx+0*t; y=tan(atan(ly/lx)*t)/tan(atan(ly/lx))*ly; label=2;}
border Cxt(t=0, 1){x=tan(atan(lx/ly)*(1-t))/tan(atan(lx/ly))*lx; y=ly+0*t; label=2;}
border Cyl(t=0, 1){x=0*t; y=A+(ly-A)*(1-t)^(2); label=2;}
border Beam(t=0, pi/2){x=A*cos(t); y=A*sin(t); label=3;}

// Mesh
real ds = 1/k/10;
int NbA = max(ceil(pi/2*A/ds),21.);
int Nbx = ceil(atan(lx/ly)/(pi/2)*(NbA))+1;
int Nby = ceil(atan(ly/lx)/(pi/2)*NbA)+1;
int NbxmA = ceil((lx-A)/lx*NbA)+1;
int NbymA = ceil((ly-A)/ly*NbA)+1;
mesh cMesh = buildmesh(Cxb(NbxmA) + Cyr(Nby) + Cxt(Nbx) + Cyl(NbymA) + Beam(-NbA));

// FE
fespace Vh(cMesh,P2);
Vh u, v;

// Resolution
solve pFourier(u,v)
  = - int2d(cMesh)(u*v*k^2 + dx(u)*dx(v) + dy(u)*dy(v))
    + int1d(cMesh,3)(y/A*v) 
    + on(1,u=0);
          
// mu computation
real mu = 0;
mu = -4*int1d(cMesh,3)(-u*y/A);
\end{lstlisting}

\section{Qualitative analysis}\label{sec.qualitative}
After solving numerically the Helmholtz problem \eqref{eq.Helmholtz.assemblies} for $\hat\gamma_x^{[1]} = 1$, the resulting forces on each rod $q$ of each assembly $n$ can be evaluated in both the $x$ and $y$ directions. From this, we can calculate the associated added masses 
$\hat\mu_x^{[n,q]}$ and $\hat\mu_y^{[n,q]}$
\begin{subequations}
\begin{align}
\hat\mu_x^{[n,q]} &= \int_{\dRod[n,q]}\hat p\cos(\theta)\,d\theta,\\
\hat\mu_y^{[n,q]} &= \int_{\dRod[n,q]}\hat p\sin(\theta)\,d\theta,    
\end{align}
\end{subequations}
with $\dRod[n,q]$ the border of the rod $q$ in assembly $n$.
Note that $\hat\mu_x^{[n,q]}$ and $\hat\mu_y^{[n,q]}$ are unchanged when $k$ is changed into $-k$ since $k$ always appears squared in the problem.

\begin{figure}
\centering
\includegraphics[width=.99\textwidth,trim=0 0cm 0 0,clip]{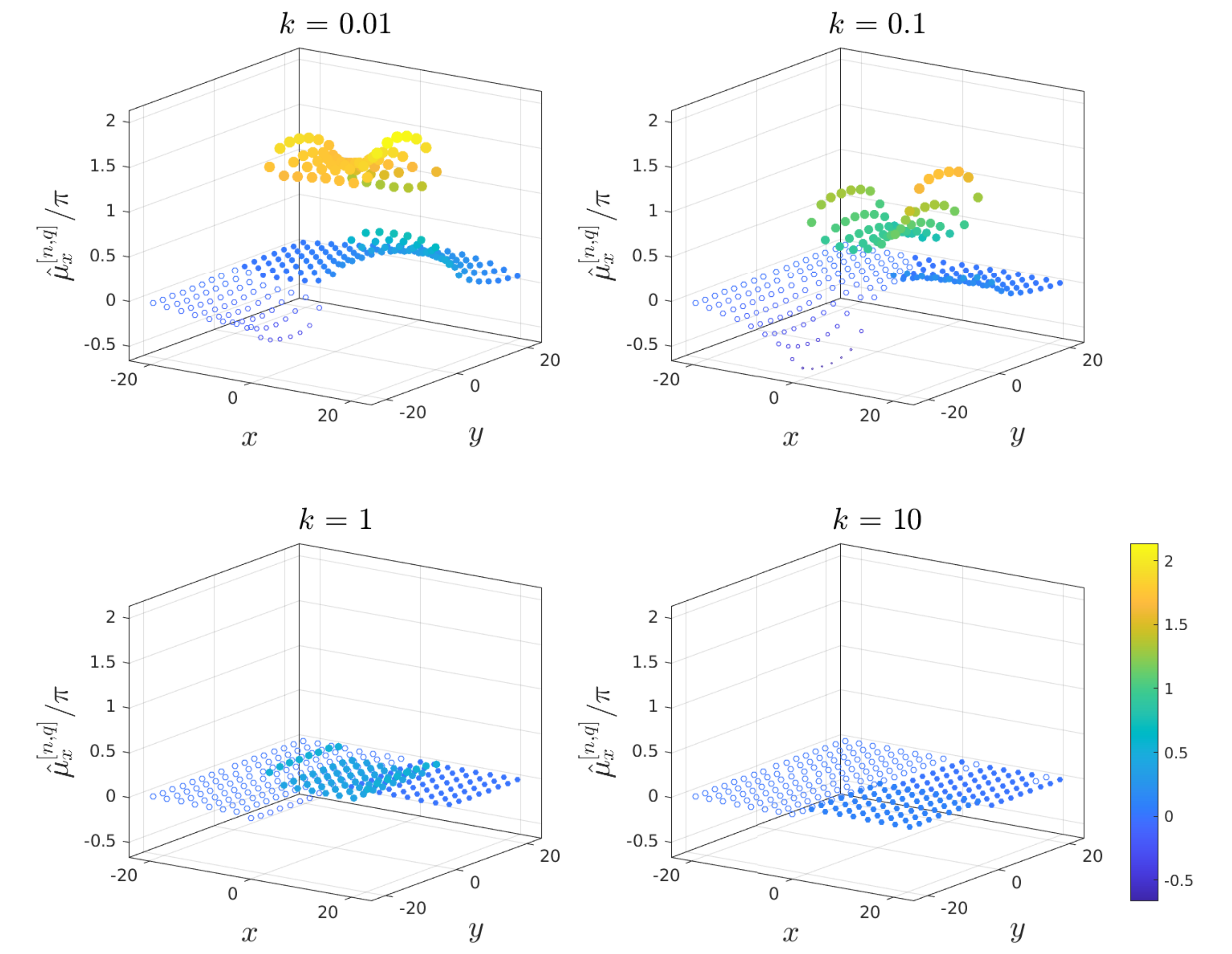}
\caption{Added mass along $x$ $\hat\mu_x^{[n,q]}$ for each rod for four values of $k$ as labelled. Full circles note positive values, hollow ones negative values. The confinement is 8~mm.}
\label{fig:4x4_8mm_rhatfx_3D}
\end{figure}

\begin{figure}
\centering
\includegraphics[width=.99\textwidth,trim=0 0cm 0 0,clip]{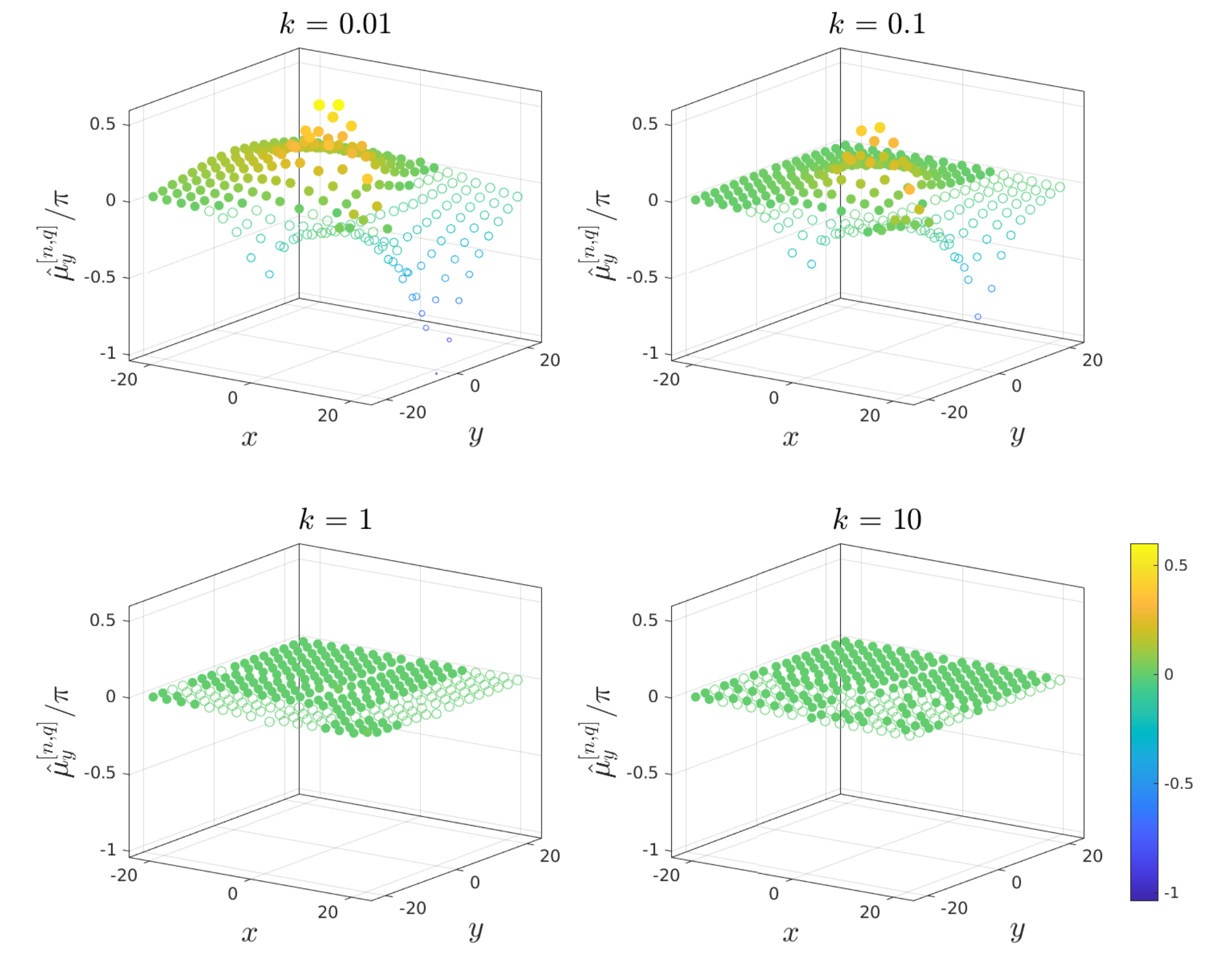}
\caption{Same as Figure~\ref{fig:4x4_8mm_rhatfx_3D} for the $y$-component of the added mass:  $\hat\mu_y^{[n,q]}$.}
\label{fig:4x4_8mm_rhatfy_3D}
\end{figure}

These added masses are illustrated in Figures~\ref{fig:4x4_8mm_rhatfx_3D} and \ref{fig:4x4_8mm_rhatfy_3D} for the 8~mm confinement. 

Analogy with the one cylinder case explains the positive values observed on assembly [1] for $\hat\mu_x^{[n,q]}$. Confinement size is about the rod diameter, that may be compared to $l_x = l_y=3$ in the previous section for which $\hat{\mu}>\pi$ when $k\ll 1$. This is consistent with the values observed for assembly [1] (even though rods are not properly confined). An interesting result is that rods in assemblies [2] to [4]  experience very small forces. As noted in \S\ref{sse::SingleCylinder_Confinement}, this phenomenon is all the more true as $k$ is large: pressure perturbations remain closer to the moving rods when $k$ is large and interactions with far away rods are almost negligible.


\section{Calculation of pressure forces in physical space}\label{app:tokeep}

\begin{figure}
\centering
\includegraphics[width=.99\textwidth,trim=0 0cm 0 0,clip]{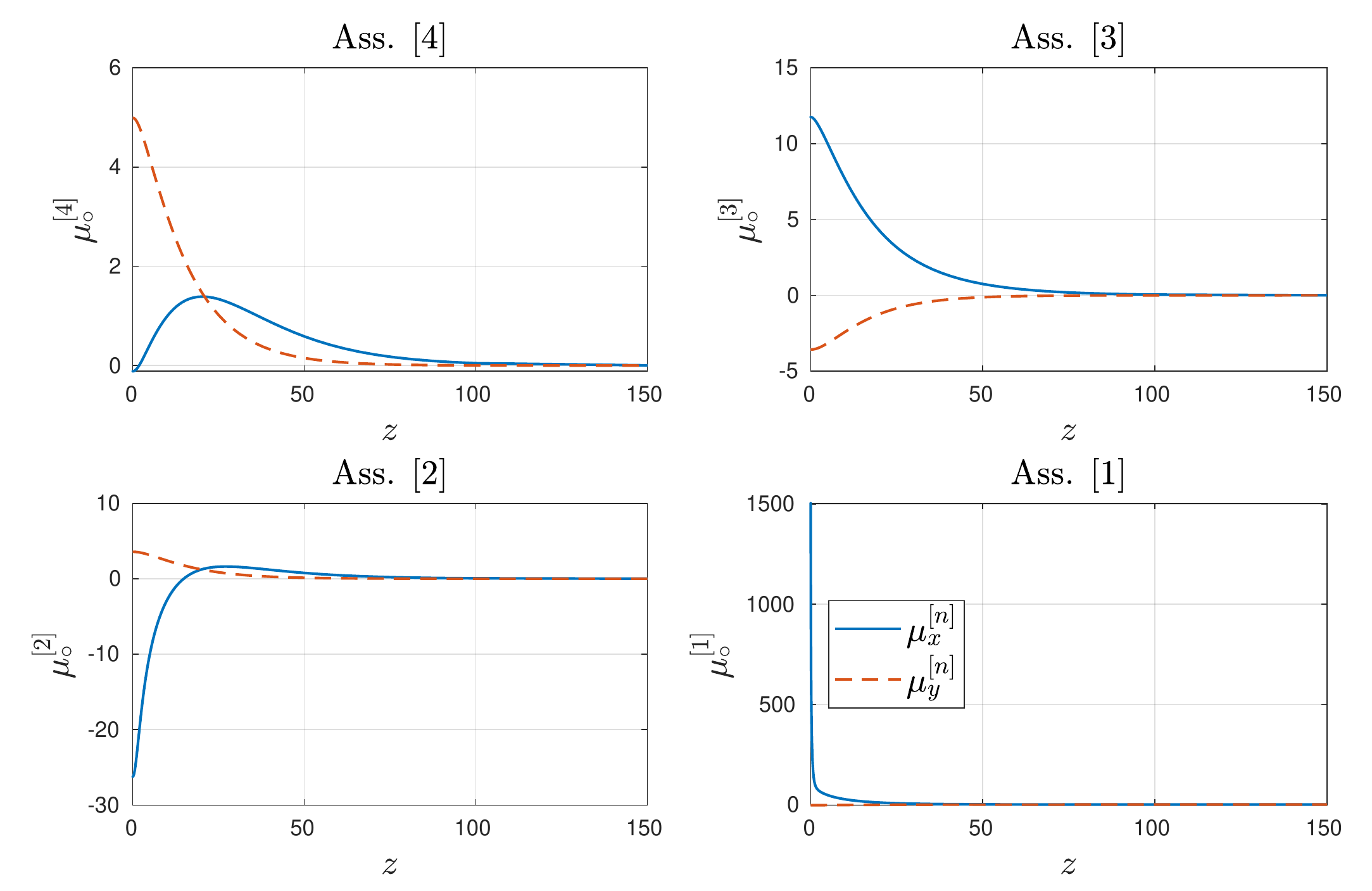}
\caption{Added mass $\mu_x^{[n]}(z)$ (blue solid line) and $\mu_y^{[n]}(z)$ (red dashed line) as a function of $z$ for the different assemblies (large confinement).}
\label{fig:mu_vs_z_assemblies}
\end{figure}

In \S\ref{sec.eqs_motion_real_space}, we explained how to compute numerically the functions  $\hat{\mu}_\circ^{[n]}(k)$, which are then interpolated with piecewise 6th-order polynomials. These interpolations can be used to compute the added masses ${\mu}_\circ^{[n]}(z)$  with an inverse Fourier transform (Figure~\ref{fig:mu_vs_z_assemblies}).

The functions ${\mu}_\circ^{[n]}(z)$ are needed in (\ref{eq.added_matrices_assemblies}\textit{a}--\textit{c}) to compute the added mass, added damping, and added stiffness matrices. 
Note that the functions $\mu_\circ^{[n]}(z)$ are non-zero on an interval $\Delta z \lesssim 100$ (figure~\ref{fig:mu_vs_z_assemblies}). 
This means that on a $z$-scale of order $l = L / A \approx 571$, these functions are almost proportional to a Dirac, such that the convolution products between $\mu_\circ^{[n]}$ and $\chi_j$, $\chi_j'$, or $\chi_j''$ appearing in (\ref{eq.added_matrices_assemblies}\textit{a}--\textit{c}) can be approximated by simple product with $\int_{-\infty}^{\infty} \mu_\circ^{[n]} \,dz$. 

As a consequence, the added mass matrix $\mathsfbi{M}_\circ^{[n]}$ and the added stiffness matrix $\mathsfbi{K}_\circ^{[n]}$ are almost diagonal (since $\langle \chi_i, \chi_j\rangle = \delta_{ij}$) and the diagonal terms of the added damping matrix $\mathsfbi{C}_\circ^{[n]}$ are almost zero (since $\langle \chi_i, \chi_i'\rangle = 0$). This can be verified by computing the first $3\times 3$ terms of the matrices $\mathsfbi{M}_x^{[n]}$ and $\mathsfbi{C}_x^{[n]}$ for $n=1$ and $4$ (we chose the assembly $[4]$ because it is the assembly for which the extension of ${\mu}_x^{[n]}(z)$ is widest, see figure~\ref{fig:mu_vs_z_assemblies})
\begin{subequations}
    \begin{eqnarray}
    \mathsfbi{M}_x^{[1]} & = &
        \left(\begin{tabular}{ccc}
            350     & 0     & 0.0182    \\
            0       & 347   & 0         \\
            0.0182  & 0     & 342
        \end{tabular}\right), \\
    \mathsfbi{M}_x^{[4]} & = &
        \left(\begin{tabular}{ccc}
            18.6    & 0     & 0.0882    \\
            0       & 17.4  & 0         \\
            0.0882  & 0     & 15.4
        \end{tabular}\right),   \\
    \mathsfbi{C}_x^{[1]} & = &
        \left(\begin{tabular}{ccc}
            0       & -3.26 & 0         \\
            3.26    & 0     & -5.79     \\
            0       & 5.79  & 0
        \end{tabular}\right), \\
    \mathsfbi{C}_x^{[4]} & = &
        \left(\begin{tabular}{ccc}
            0       & -0.167  & 0       \\
            0.167   & 0       & -0.272  \\
            0       & 0.272   & 0
        \end{tabular}\right).
    \end{eqnarray}
\end{subequations}

Note that the matrices $\mathsfbi{K}_\circ^{[n]}$ satisfy $(\mathsfbi{K}_\circ^{[n]})_{ij} = -k_j^2 (\mathsfbi{M}_\circ^{[n]})_{ij}$, such that they share the same characteristic of being almost diagonal with the matrices $\mathsfbi{M}_\circ^{[n]}$.

\end{document}